\documentclass[a4paper,12pt]{article}
\usepackage[T1]{fontenc}
\usepackage[utf8]{inputenc}
\usepackage{lmodern}
\usepackage{graphicx}
\usepackage{epstopdf}
\usepackage{color}
\usepackage{bbold}
\usepackage{amsmath}
\usepackage{array}
\usepackage{booktabs}

\newcommand{\DimUV}{\mathrm{Dim\,UV}}

\title{Spontaneous symmetry breaking as a mechanism for the Gribov scale in a BRST-invariant formulation}

\author{
M.A.L.Capri$^{a}$\footnote{email: caprimarcio@gmail.com},V. E. R. Lemes$^{a}$\footnote{email: verlemes@gmail.com}
\\
\small \em $^a$Instituto de F\'\i sica, Universidade do Estado do Rio de
Janeiro,\\
\small \em Rua S\~{a}o Francisco Xavier 524, Maracan\~{a}, Rio de Janeiro - RJ,
20550-013, Brazil\\
}

\begin{document}
\maketitle
\begin{abstract}
We propose a BRST-invariant formulation of the Gribov--Zwanziger model in the Landau gauge, in which the Gribov scale is produced by the spontaneous breaking of a scalar quartet symmetry, rather than through the explicit breaking term of the conventional construction. We show that the resulting action preserves the ghost equation of the Landau gauge and reproduces the characteristic Gribov-type gauge propagator of the original Gribov scenario, a BRST-invariant mass term of the refined type being accommodated by the same mechanism. We establish the renormalizability of the model to all orders by means of algebraic renormalization techniques, no counterterm being required outside those already present in the classical action. We further show that the condensation of the quartet leaves unchanged the cohomology through which the observables are characterized, the operator realized in the displaced vacuum having the same cohomology as that of the symmetric phase, so that the mass scale is produced at no cost in the physical content of the theory, and the construction of genuine gauge-invariant observables in the Gribov--Zwanziger setting is thereby left open to the usual cohomological methods.
\end{abstract}

\section{Introduction}
The quantization of Yang--Mills theories in the Landau gauge leads to the well-known Gribov ambiguity, i.e., the fact that the gauge-fixing condition does not uniquely select a gauge orbit \cite{Gribov:1977wm}. A prominent way to deal with this issue is the Gribov--Zwanziger construction, which restricts the functional integral to the so-called Gribov region, thereby dynamically generating a mass scale known as the Gribov scale \cite{Zwanziger:1991ac,Zwanziger:1992qr}. In fact, this construction captures important infrared features of the theory, such as the suppression of the transverse gauge propagator and the enhancement of the ghost propagator, features that are believed to be relevant for the confinement scenario. For a comprehensive review of the Gribov--Zwanziger framework and its implications for QCD dynamics, we refer to \cite{Vandersickel:2012tz}.

However, the standard implementation of the Gribov--Zwanziger framework explicitly breaks the BRST symmetry. This breaking poses a genuine difficulty, since physical observables in a gauge theory are typically identified with BRST-invariant quantities, and the very introduction of the Gribov scale, through a non-local horizon condition, effectively amounts to an explicit symmetry-breaking term in the action. Recent developments have shown that a nilpotent BRST symmetry can, in fact, be preserved in the Gribov--Zwanziger setting through suitable modifications of the gauge-fixing procedure \cite{Capri:2015ixa}.

Given that the Gribov scenario demands the introduction of a mass scale, a natural alternative is to generate this scale via spontaneous symmetry breaking, where a scalar field acquires a non-trivial vacuum expectation value. Put in another way, the scale would then appear dynamically, without the need for any explicit breaking of the BRST invariance. This possibility is further suggested by the observation that gauge fields do not present a Gribov problem along the directions broken by a spontaneous symmetry breaking mechanism \cite{Amaral:2024}, which points to confinement itself as possibly resulting from a hidden breaking of this kind. This observation motivates the present work: our aim is to construct a Gribov--Zwanziger action that preserves the BRST symmetry exactly, relying instead on spontaneous symmetry breaking to generate the Gribov scale. In a related context, the spontaneous breaking of the BRST symmetry itself in the presence of the Gribov horizon has also been explored \cite{Dudal:2012sb}, although our approach differs in an essential way. In the present construction the symmetry is not lost when the quartet fields acquire their vacuum expectation values: the invariance of the action is exact in both phases, and what the displacement of the vacuum modifies is the way in which the symmetry is realized, the observables continuing to be characterized by the cohomology of the generator that is realized in the chosen vacuum \cite{Amaral:2022,Amaral:2023}. We return to this point in Section~\ref{sec:brokenWard}, where we show that in the present model this cohomology is in fact the same one in both phases.

In this paper, we present a BRST-invariant formulation of the Gribov--Zwanziger model in the Landau gauge. Our strategy is to introduce a scalar quartet that condenses, providing a non-vanishing vacuum expectation value that plays the role of the Gribov parameter. We show that the resulting action maintains the crucial algebraic properties of the Landau gauge, in particular the ghost equation, while yielding the characteristic Gribov-type propagators. The model is shown to be renormalizable by means of algebraic renormalization techniques \cite{Capri:2011wp}, the original Gribov scenario being obtained directly and a mass term of the refined type being accommodated as a particular case. For additional details on the renormalization of the Gribov--Zwanziger action in the presence of modified BRST symmetries, we refer to \cite{Capri:2011wp,Dudal:2010fq}.

Our results may be summarized as follows. The action we construct is local and invariant under an exact nilpotent BRST symmetry, and it preserves the ghost equation of the Landau gauge, so that the ghost does not renormalize independently and its propagator retains the form $1/p^{2}$. Once the scalar quartet condenses, the gauge field acquires the Gribov propagator, with complex conjugate poles at $p^{2}=\pm i\gamma^{2}$, the scale being fixed by the vacuum of the scalar sector together with the gauge coupling, with which it vanishes; the same condensate accommodates, for non-vanishing $a_{7}$, a BRST-invariant gauge mass of the kind employed in the refined scenario, and the mixed propagators between the gauge field and the auxiliary fields fall off as $\sigma^{2}/p^{4}$, improving the convergence of the loop integrals. The model is renormalizable to all orders by the algebraic method, with no counterterm beyond those already present in the classical action. Finally, and this is the point on which the whole construction rests, the condensation of the quartet does not alter the cohomology through which the observables are defined: the operator realized in the displaced vacuum is nilpotent, is an exact invariance of the action, and has the same cohomology as the one of the symmetric phase.

In the sequence of this work, we proceed as follows. In Section 2 we briefly review Yang--Mills theory in the Landau gauge and recall the relevant BRST transformations and algebraic identities. In Section 3 we introduce the auxiliary fields and the scalar quartet, together with the global $U(1)$ symmetry that controls their interactions, present the complete BRST-invariant action, and discuss the spontaneous symmetry breaking mechanism. In Section 4 we present the tree-level propagators of the model, where the Gribov-type behavior is explicitly derived. In Section 5 we analyze the renormalization properties of the theory, establishing its algebraic renormalizability through a set of functional identities. In Section 6 we discuss the broken phase, namely the form of the Ward identities once the vacuum is displaced, the realization of the BRST symmetry in the non-trivial vacuum, and the one-loop vacuum condition. In Section 7 we discuss in what sense the mechanism is sufficient to reproduce the bilinear content of the Gribov condition. Section 8 contains our conclusions, together with the points that remain open. The details of the solution of the renormalization system are collected in Appendix~\ref{app:renorm}.

\section{Yang-Mills in the Landau gauge}
The Yang-Mills action is (Euclidean)
\begin{eqnarray}
S_{YM}&=& \int d^{4}x \{ \frac{1}{4}F^{a\mu\nu}F^{a}_{\mu\nu}\}\nonumber \\
F^{a}_{\mu\nu}&=& \partial_{\mu}A^{a}_{\nu}- \partial_{\nu}A^{a}_{\mu}+gf^{abc}A^{b}_{\mu}A^{c}_{\nu}\nonumber \\
sS_{YM}&=&0\nonumber \\
sA_{\mu }^{a}&=& -(\partial _{\mu }c^{a}+gf^{abc}A_{\mu }^{b}c^{c})
\end{eqnarray}
The Landau gauge is basically obtained from the BRST variation of $\frac{1}{2} A^{a}_{\mu}A^{a\mu}$, as follows:
\begin{eqnarray}
s \int d^{4}x\{\frac{1}{2} A^{a}_{\mu}A^{a\mu}\} &=& \int d^{4}x \{ c^{a}\partial_{\mu}A^{a\mu}\}\nonumber \\
s \partial_{\mu}A^{a\mu}&=& -\partial_{\mu}(D^{\mu}c)^{a}\nonumber \\
sc^{a}&=&\frac{g}{2}f^{abc}c^{b}c^{c}\nonumber \\
\end{eqnarray}

Thus we obtain the Landau gauge condition,
\begin{eqnarray}
s\{ \overline{c}^{a} \partial_{\mu}A^{a\mu} \} &=& ib^{a}\partial_{\mu}A^{a\mu} + \overline{c}^{a} \partial_{\mu}(D^{\mu}c)^{a}\nonumber \\
s\overline{c}^{a}&=& i b^{a}, \,\,\, sb^{a}=0
\end{eqnarray}
But what matters, from an algebraic point of view, is to have integrated or non-integrated Ward identities that characterize the gauge. It is not difficult to see that the following equation holds in this gauge,
\begin{eqnarray}
S&=&\int d^{4}x \{ \frac{1}{4}F^{a\mu\nu}F^{a}_{\mu\nu} + ib^{a}\partial_{\mu}A^{a\mu} + \overline{c}^{a} \partial_{\mu}(D^{\mu}c)^{a}\}\nonumber \\
G^{a}(S)&=& \int d^{4}x \{ \frac{\delta S}{\delta c^{a} } -igf^{abc} \overline{c}^{b}\frac{\delta S}{\delta b^{c} } \} = 0
\end{eqnarray}
and it characterizes the Landau gauge together with the equation
\begin{eqnarray}
\frac{\delta S}{\delta b^{a} }&=& i\partial_{\mu}A^{a\mu}\nonumber \\
\end{eqnarray}

Thus, constructing the Gribov restriction in the Landau gauge must preserve this equation, or at least the properties arising from it, namely that the ghost does not renormalize independently.

\section{Construction of the BRST-invariant action}

The requirement of preserving the Landau gauge ghost equation — specifically, the property that the ghost does not renormalize independently — implies that the localizing fields introduced for the Gribov construction must carry a part of their BRST transformation that behaves as matter, in addition to the usual quartet structure. It is also important to observe that, in the original Gribov-Zwanziger formulation, fields with two adjoint indices are used, coupling to a group structure $f^{abc}B^{bc}_{\mu}$, such that the field is antisymmetric in the group indices, or at least only its antisymmetric part is relevant. Bearing in mind that the fields always appear coupled in this format, we will simply introduce fields with a direct group index, that is, something like $f^{abc}B^{bc}_{\mu}\rightarrow\varphi^{a}_{\mu}$.

We thus introduce the following fields, with the symmetries
\begin{eqnarray}
\label{symm}
s\overline{\omega }_{\mu }^{a} &=& \overline{\varphi}_{\mu }^{a} + gf^{abc}c^{b}\overline{\omega }_{\mu }^{c}\nonumber \\
s\overline{\varphi}_{\mu }^{a} &=& gf^{abc}c^{b}\overline{\varphi}_{\mu }^{c}\nonumber \\
s\varphi_{\mu }^{a} &=& \omega _{\mu }^{a}+gf^{abc}c^{b}\varphi_{\mu }^{c} \nonumber \\
s\omega _{\mu }^{a} &=& gf^{abc}c^{b}\omega _{\mu }^{c}
\end{eqnarray}
It is trivial to observe that
\begin{eqnarray}
s\{ (D_{\nu}\varphi^{\mu})^{a}\} &=& (D_{\nu}\omega^{\mu})^{a} + gf^{abc}c^{b}(D_{\nu}\varphi^{\mu})^{c}\nonumber \\
s\{\overline{\omega}^{a}_{\mu} \partial^{\nu} (D_{\nu}\varphi^{\mu})^{a}\} &=& \overline{\varphi}^{a}_{\mu} \partial^{\nu} (D_{\nu}\varphi^{\mu})^{a} - \overline{\omega}^{a}_{\mu} \partial^{\nu} (D_{\nu}\omega^{\mu})^{a} - gf^{abc} \overline{\omega}^{a}_{\mu}(\partial^{\nu}c^{b}) (D_{\nu}\varphi^{\mu})^{c}
\end{eqnarray}
This result is only possible because the Gribov localizing fields carry a matter-type transformation part. It is also relevant to observe that the ghost appears explicitly with a derivative: since we want the ghost equation to be obeyed, this specific form only arises because part of the BRST transformation of the fields in the quartet is of the matter type, and without this term underived ghosts would appear. In fact, it is precisely the quartet structure of the fields that guarantees a trivial BRST action in the trivial vacuum.

Now we introduce a simple scalar quartet
\begin{eqnarray}
\label{quartet}
s\overline{\psi } &=& \overline{\phi } \nonumber \\
s\overline{\phi } &=& 0 \nonumber \\
s\phi &=& \psi \nonumber \\
s\psi &=& 0.
\end{eqnarray}
A few remarks on the nature of the scalar quartet fields are in order. In the trivial vacuum, where $\langle \phi \rangle = \langle \bar{\phi} \rangle = 0$, these fields form a BRST quartet and therefore decouple from the physical spectrum: in this phase, they do not contribute to the dynamics and can be integrated out without altering the infrared behavior of the theory. However, in the non-trivial vacuum with $\langle \phi \rangle = \langle \bar{\phi} \rangle = \nu \neq 0$, the situation changes drastically, since the scalar fields acquire non-vanishing expectation values, which induces mixing with the gauge and ghost sectors; consequently, they become dynamically relevant and their renormalization must be carefully accounted for. Thus, while the quartet fields are auxiliary in the symmetric phase, they become physical, in the sense that they participate in the quantum dynamics and affect the renormalization of the theory, once spontaneous symmetry breaking takes place. This is analogous to the standard Faddeev--Popov ghost system, which also forms a BRST quartet in the trivial vacuum but becomes physically relevant through its coupling to the gauge fields.

Our goal is to introduce an interaction term with the gauge field that preserves the equations defining the Landau gauge, and to this end we take the scalar quartet as consisting of fields of ultraviolet dimension one. To control the allowed interactions between the scalar and vector quartets, a global $U(1)$ symmetry is imposed, necessary to prevent the appearance of quartic terms that would otherwise affect the Gribov construction; this set of symmetries turns out to be sufficient to fix the mixed terms in the action without ambiguity.

The most obvious solution for interaction terms between the quartets and the gauge field consists of a product of UV dimension 3, which is in turn fixed by the same global $U(1)$ symmetry. Thus,
\begin{eqnarray}
s\{ \overline{\psi }\varphi^{a}_{\mu}A^{a\mu}\}&=& (\overline{\phi}\varphi^{a}_{\mu}A^{a\mu}- \overline{\psi }\omega^{a}_{\mu}A^{a\mu} + \overline{\psi}\varphi^{a}_{\mu}\partial^{\mu}c^{a})\nonumber \\
s\{ \phi \overline{\omega}^{a}_{\mu}A^{a\mu}\}&=& (\psi\overline{\omega}^{a}_{\mu}A^{a\mu}+\phi\overline{\varphi}^{a}_{\mu}A^{a\mu}+\phi\overline{\omega}^{a}_{\mu}\partial^{\mu}c^{a}).
\end{eqnarray}

These terms have UV dimension 3, but simply take the form of a field with a bar followed by one without a bar, which is exactly what characterizes a global symmetry. Before writing them into the action, we fix the normalization of their coefficients. We shall take them of the form $a_{i}gm$, with the gauge coupling appearing explicitly and $a_{1}$, $a_{2}$ dimensionless. The reason is that these are the terms which, once the scalars condense, mix the gauge field with the auxiliary fields and thereby generate the Gribov scale, according to $\gamma^{4}=2a_{1}a_{2}g^{2}m^{2}\nu^{2}$; coefficients independent of the coupling would leave the gauge propagator of Gribov type in the limit $g\to0$, in which no gauge copies are present, whereas with the normalization adopted here the scale vanishes together with the coupling, as it must if it is to originate in the restriction of the functional integral. The consequences of this normalization for the renormalization of the model are discussed in Section 5, and those for the vacuum condition in Section~\ref{sec:brokenWard}. We thus have
\begin{eqnarray}
\overline{\psi }&&\rightarrow e^{i\theta}\overline{\psi }\nonumber \\
\varphi^{a}_{\mu}&&\rightarrow e^{-i\theta}\varphi^{a}_{\mu}\nonumber \\
\psi &&\rightarrow e^{-i\theta}\psi\nonumber \\
\overline{\omega}^{a}_{\mu}&&\rightarrow e^{i\theta}\overline{\omega}^{a}_{\mu}\nonumber \\
\overline{\phi}&&\rightarrow e^{i\theta}\overline{\phi}\nonumber \\
\phi&&\rightarrow e^{-i\theta}\phi\nonumber \\
\overline{\varphi}^{a}_{\mu}&&\rightarrow e^{i\theta}\overline{\varphi}^{a}_{\mu}\nonumber \\
\omega^{a}_{\mu}&&\rightarrow e^{-i\theta}\omega^{a}_{\mu}
\label{U}
\end{eqnarray}
This global symmetry is precisely what prevents the quartic terms that would spoil the Gribov construction, and it is sufficient to fix the mixed terms in the action without further ambiguity.

The resulting action is then

\begin{eqnarray}
S&=&\int d^{4}x \{ \frac{1}{4}F^{a\mu\nu}F^{a}_{\mu\nu} + ib^{a}\partial_{\mu}A^{a\mu} + \overline{c}^{a} \partial_{\mu}(D^{\mu}c)^{a}\nonumber \\
&+& \overline{\varphi}^{a}_{\mu} \partial^{\nu} (D_{\nu}\varphi^{\mu})^{a} - \overline{\omega}^{a}_{\mu} \partial^{\nu} (D_{\nu}\omega^{\mu})^{a} - gf^{abc}\overline{\omega}^{a}_{\mu}(\partial^{\nu}c^{b}) (D_{\nu}\varphi^{\mu})^{c}\nonumber \\
&+& a_{1}gm (\overline{\phi}\varphi^{a}_{\mu}A^{a\mu}- \overline{\psi }\omega^{a}_{\mu}A^{a\mu} + \overline{\psi}\varphi^{a}_{\mu}\partial^{\mu}c^{a})\nonumber \\
&+& a_{2}gm (\psi\overline{\omega}^{a}_{\mu}A^{a\mu}+\phi\overline{\varphi}^{a}_{\mu}A^{a\mu}+\phi\overline{\omega}^{a}_{\mu}\partial^{\mu}c^{a})\nonumber \\
&+& \partial^{\mu}\overline{\phi}\partial_{\mu}\phi - \partial^{\mu}\overline{\psi}\partial_{\mu}\psi - m^{2}(\overline{\phi }\phi - \overline{\psi }\psi ) + \frac{\lambda}{2} (\overline{\phi }\phi - \overline{\psi }\psi )^{2} \}
\end{eqnarray}

The scalar sector includes a potential that has a non-trivial minimum. This vacuum expectation value spontaneously breaks the global $U(1)$ symmetry while preserving the BRST symmetry. The operator algebra itself is not modified by the vacuum choice; the shift by $\nu$ only generates one-point Ward identities whose constant terms are fixed by the vacuum condition, as detailed in Section~\ref{sec:brokenWard}. It should be stressed, at this point, that the displacement of the vacuum entails no loss of the BRST symmetry. The invariance of the action is exact in both phases, and what the non-trivial vacuum modifies is the realization of the symmetry, the relevant generator in the broken phase being the vacuum operator obtained by displacing the fields in the original transformations \cite{Amaral:2022,Amaral:2023}, with respect to which the observables of the broken phase are to be classified. We show in Section~\ref{sec:brokenWard} that, for the quartet employed here, this classification leads to the same cohomology in both phases. Consequently, a mass scale is introduced into the theory in a BRST-invariant manner, circumventing the explicit symmetry breaking typical of the conventional Gribov--Zwanziger approach.

It is possible to observe that the potential in the scalar quartet sector has a non-trivial minimum, such that
\begin{eqnarray}
<\overline{\phi}><\phi> &=& \frac{m^{2}}{\lambda}= \nu^{2}
\end{eqnarray}
Let us now present these symmetries in functional form. First, we have the symmetry defined in (\ref{U}),

\begin{eqnarray}\label{U1}
U(S)&=& \int d^{4}x \{ \overline{\varphi}^{a}_{\mu}\frac{\delta S}{\delta \overline{\varphi}^{a}_{\mu}}+ \overline{\omega}^{a}_{\mu}\frac{\delta S}{\delta \overline{\omega}^{a}_{\mu}} + \overline{\psi}\frac{\delta S}{\delta \overline{\psi}}+ \overline{\phi}\frac{\delta S}{\delta \overline{\phi}}\nonumber \\
&-&\varphi^{a}_{\mu}\frac{\delta S}{\delta \varphi^{a}_{\mu}} -\omega^{a}_{\mu}\frac{\delta S}{\delta \omega^{a}_{\mu}}- \psi\frac{\delta S}{\delta \psi} - \phi\frac{\delta S}{\delta \phi}\} = 0
\end{eqnarray}
and
\begin{eqnarray}
X(S)&=&\int d^{4}x \{ \varphi^{a}_{\mu}\frac{\delta S}{\delta \omega^{a}_{\mu}}-\overline{\omega}^{a}_{\mu}\frac{\delta S}{\delta \overline{\varphi}^{a}_{\mu}} - \overline{\psi}\frac{\delta S}{\delta \overline{\phi}}+\phi\frac{\delta S}{\delta \psi}\} = 0\label{XS}
\end{eqnarray}

The $X$-symmetry carries ghost number $-1$ and satisfies the commutation relation $\{X, \beta\} = -U$, showing that the $U(1)$ symmetry emerges, in fact, from the anticommutator of $X$ with the BRST operator; consequently, preserving BRST automatically enforces the $U(1)$ invariance of the renormalized theory. In the broken phase, the same relation implies that the terms proportional to $\nu$ are tadpole contributions to the Ward identities, rather than new symmetry generators; see Section~\ref{sec:brokenWard}.

%Basta depois introduzir as fontes para extender para a ação $\Sigma$. Neste caso usando o operador $\beta_{\Sigma}$ temos que
%$\{X,\beta_{\Sigma}\}=-U$

The term
\begin{eqnarray}
s\{ \overline{\psi}\phi\frac{1}{2} A^{a\mu}A^{a}_{\mu}\}&=& (\overline{\phi }\phi - \overline{\psi }\psi )\frac{1}{2} A^{a\mu}A^{a}_{\mu} + \overline{\psi}\phi A^{a\mu}\partial_{\mu}c^{a}\label{refined}
\end{eqnarray}
does not break the ghost equation and is invariant under (\ref{XS}), and can therefore be placed directly into the action. When the coefficient $a_7$ is non-zero, this term corresponds precisely to the ``Refined Gribov--Zwanziger'' model, now realized in a fully BRST-invariant manner through spontaneous symmetry breaking.

As an example of the constraints at work in building the action, we note that a term of the form
\begin{equation}
s\{ \nu^{2} \overline{\omega}^{a}_{\mu}A^{a\mu}\}= \nu^{2} \overline{\varphi}^{a}_{\mu}A^{a\mu} - \nu^{2}\overline{\omega}^{a}_{\mu}\partial^{\mu}c^{a}
\end{equation}
does not satisfy the symmetries (\ref{U1}, \ref{XS}), and therefore cannot be introduced into the action. The complete action is then

\begin{eqnarray}
S&=&\int d^{4}x \{ \frac{1}{4}F^{a\mu\nu}F^{a}_{\mu\nu} + ib^{a}\partial_{\mu}A^{a\mu} + \overline{c}^{a} \partial_{\mu}(D^{\mu}c)^{a}\nonumber \\
&+& \overline{\varphi}^{a}_{\mu} \partial^{\nu} (D_{\nu}\varphi^{\mu})^{a} - \overline{\omega}^{a}_{\mu} \partial^{\nu} (D_{\nu}\omega^{\mu})^{a} - gf^{abc}\overline{\omega}^{a}_{\mu}(\partial^{\nu}c^{b}) (D_{\nu}\varphi^{\mu})^{c}\nonumber \\
&+& a_{1}gm (\overline{\phi}\varphi^{a}_{\mu}A^{a\mu}- \overline{\psi }\omega^{a}_{\mu}A^{a\mu} + \overline{\psi}\varphi^{a}_{\mu}\partial^{\mu}c^{a})\nonumber \\
&+& a_{2}gm (\psi\overline{\omega}^{a}_{\mu}A^{a\mu}+\phi\overline{\varphi}^{a}_{\mu}A^{a\mu}+\phi\overline{\omega}^{a}_{\mu}\partial^{\mu}c^{a})\nonumber \\
&+& a_{7}\{(\overline{\phi }\phi - \overline{\psi }\psi )\frac{1}{2} A^{a\mu}A^{a}_{\mu} + \overline{\psi}\phi A^{a\mu}\partial_{\mu}c^{a}\} \nonumber \\
&+& \partial^{\mu}\overline{\phi}\partial_{\mu}\phi - \partial^{\mu}\overline{\psi}\partial_{\mu}\psi - m^{2}(\overline{\phi }\phi - \overline{\psi }\psi ) + \frac{\lambda}{2} (\overline{\phi }\phi - \overline{\psi }\psi )^{2} \}
\end{eqnarray}

\section{Free propagators of the model}

In this section we present the free propagators of the model, obtained from the bilinear expansion of the action around the vacuum. In Euclidean space, we adopt the Fourier transform such that $\partial_\mu \to -i p_\mu$, so that $\partial^2 \to -p^2$, with $p^2 = p_\mu p_\mu \ge 0$. The non-zero expectation values are given by $\langle \phi \rangle = \langle \overline{\phi} \rangle = \nu$, with $\nu^2 = m^2/\lambda$, and we define the parameters

\begin{eqnarray}
\gamma^4 = 2a_{1}a_{2}g^{2}m^2 \nu^2 = 2a_{1}a_{2}\sigma^{4}, \qquad \sigma^2 = gm \nu.
\end{eqnarray}
The dynamical determination of $\nu$, and hence of $\gamma$, follows from the broken-phase Ward identities discussed in Section~\ref{sec:brokenWard}.

The fluctuating fields are then defined as

\begin{eqnarray}
\phi = \nu + \frac{\phi_1 + i\phi_2}{\sqrt{2}}, \quad
\overline{\phi} = \nu + \frac{\phi_1 - i\phi_2}{\sqrt{2}}, \quad
\psi = \frac{\psi_1 + i\psi_2}{\sqrt{2}}, \quad
\overline{\psi} = \frac{\psi_1 - i\psi_2}{\sqrt{2}},
\end{eqnarray}
where it should be kept in mind that $\overline{\phi}$ is not the complex conjugate of $\phi$, as is evident from the transformations (\ref{quartet}), so that this decomposition is to be read as a convenient parametrization of four independent fields rather than as a statement of conjugation.

\subsection{Scalar sector}

At the quadratic order the scalar sector does not mix with the gauge field nor with the auxiliary vector fields, since the interaction terms of the action become trilinear once a single field is replaced by its expectation value. The scalar propagators therefore follow from the potential alone. Using the vacuum condition $\lambda\nu^{2}=m^{2}$, the term linear in the fluctuations cancels and the quadratic part of the potential reduces to
\begin{eqnarray}
V_{2}&=& \frac{m^{2}}{2}\left(\frac{\phi_1 + i\phi_2}{\sqrt{2}} + \frac{\phi_1 - i\phi_2}{\sqrt{2}}\right)^{2}= m^{2}\phi_{1}^{2},
\end{eqnarray}
so that only the combination $\phi_{1}$ acquires a mass, while no mass term at all survives for $\phi_{2}$ and for $\psi$ and $\overline{\psi}$. The propagators are then
\begin{eqnarray}
\langle \phi_1(p) \phi_1(-p) \rangle &=& \frac{1}{p^2 + 2m^2},\,\,
\langle \phi_2(p) \phi_2(-p) \rangle = \frac{1}{p^2},\nonumber \\
\langle \psi_i(p) \psi_i(-p) \rangle &=&- \frac{1}{p^2}, \ \quad i = 1,2.
\end{eqnarray}

Two features of this spectrum deserve a comment. In the first place, $\phi_{2}$ is the direction along which the global $U(1)$ acts on the vacuum, since an infinitesimal transformation (\ref{U}) gives $\delta\phi_{2}\propto\nu$, and it is massless, as it must be for the would-be Goldstone mode of the spontaneously broken $U(1)$. In the second place, the propagator of $\psi_i$ carries a negative sign relative to that of $\phi_2$, which is characteristic of a quartet system, in which $\psi$ acts as a ghost, and the fields $\psi$ and $\overline{\psi}$ are massless as well. It should be observed, however, that the massive mode $\phi_{1}$ is not degenerate with the field into which the BRST operator maps it, a point to which we return in Section~\ref{sec:cohomology}.

\subsection{Gauge and auxiliary vector sector}

In the Landau gauge, the gauge propagator, projected onto its transverse part, is given by

\begin{eqnarray}
\langle A_\mu^a(p) A_\nu^b(-p) \rangle =- \delta^{ab} \, \frac{p^2}{p^4 + \gamma^4} \, P_{\mu\nu}(p),
\end{eqnarray}

where $P_{\mu\nu}(p) = \delta_{\mu\nu} - p_\mu p_\nu/p^2$ is the transverse projector. From the bilinear expansion around the non-trivial vacuum, this transverse gauge propagator exhibits the characteristic Gribov-type behavior, with poles at the complex momenta $p^{2} = \pm i\gamma^{2}$. It is worth observing that this is the gauge propagator result for $a_7$ equal to zero, which corresponds to the original Gribov situation; the case of non-vanishing $a_7$, in which the term (\ref{refined}) contributes a mass to the gauge field, is treated below.

The mixed propagators between the gauge field and the Gribov fields $\varphi$ and $\overline{\varphi}$ are

\begin{eqnarray}
\langle A_\mu^a(p) \varphi_\nu^b(-p) \rangle &=& -\delta^{ab} \, \frac{a_{2}\sigma^2}{p^4 + \gamma^4} \, P_{\mu\nu}(p)\nonumber \\
\langle A_\mu^a(p) \overline{\varphi}_\nu^b(-p) \rangle &=& -\delta^{ab} \, \frac{a_{1}\sigma^2}{p^4 + \gamma^4} \, P_{\mu\nu}(p).
\end{eqnarray}
It is worth observing that the coefficient of $\langle A\varphi\rangle$ is $a_{2}$ and that of $\langle A\overline{\varphi}\rangle$ is $a_{1}$, and not the other way around, since the field $\varphi$ reaches the gauge field only through its propagation into $\overline{\varphi}$, which is the leg that couples with the strength $a_{2}\sigma^{2}$. The transverse projector appears because the gauge field is transverse in the Landau gauge, so that any correlation function containing $A_{\mu}$ is transverse in that index.

These mixed propagators exhibit an improved ultraviolet behavior. In fact, they fall off as $\sigma^{2}/p^{4}$ in the ultraviolet, and therefore, from the point of view of loop integrals, contribute to making the integrals considerably more convergent.

The propagator between $\varphi$ and $\overline{\varphi}$ receives, besides its free part, the contribution of the gauge field exchanged between the two mixing terms, and is given by

\begin{eqnarray}
\langle \varphi_\mu^a(p) \overline{\varphi}_\nu^b(-p) \rangle &=& \delta^{ab}\left\{ \frac{1}{p^{2}}\left[1-\frac{a_{1}a_{2}\sigma^{4}}{p^{4}+\gamma^{4}}\right]P_{\mu\nu}(p) + \frac{1}{p^{2}}L_{\mu\nu}(p)\right\}\nonumber \\
&=& \delta^{ab}\left\{ \frac{p^{4}+\frac{1}{2}\gamma^{4}}{p^{2}\left(p^{4}+\gamma^{4}\right)}P_{\mu\nu}(p) + \frac{1}{p^{2}}L_{\mu\nu}(p)\right\},
\end{eqnarray}
with $L_{\mu\nu}(p)=p_{\mu}p_{\nu}/p^{2}$ the longitudinal projector, the longitudinal part being unaffected by the mixing because the gauge field has no longitudinal component in the Landau gauge. We note that this propagator retains the massless pole $1/p^{2}$, as it does in the standard Gribov--Zwanziger formulation, the effect of the condensate being confined to the transverse part.

\subsection{Ghost sector}

The bilinear action for the ghosts, including the mixing term coming from the VEV $\langle \phi \rangle = \nu$, is

\begin{eqnarray}
S_{\text{ghost}} = \int d^4x \left[ \overline{c}^a \partial^2 c^a - \overline{\omega}_\mu^a \partial^2 \omega^{a\mu} + a_{2}\sigma^2 \, \overline{\omega}_\mu^a \partial^\mu c^a \right].
\end{eqnarray}

In momentum space, using $\partial_\mu \to -i p_\mu$ and $\partial^2 \to -p^2$, we obtain

\begin{eqnarray}
S_{\text{ghost}} = \int \frac{d^4p}{(2\pi)^4} \left[ -p^2 \overline{c} c + p^2 \overline{\omega}_\mu \omega^\mu - i a_{2}\sigma^2 p^\mu \overline{\omega}_\mu c \right].
\end{eqnarray}

The resulting propagators are

\begin{eqnarray}
\langle c^a(p) \overline{c}^b(-p) \rangle &=& \delta^{ab} \, \frac{1}{p^2},\nonumber \\
\langle \omega_\mu^a(p) \overline{\omega}_\nu^b(-p) \rangle &=& -\delta^{ab} \, \frac{\delta_{\mu\nu}}{p^2},\nonumber \\
\langle \omega_\mu^a(p)\overline{c}^b(-p) \rangle &=& \delta^{ab} \, \frac{i a_{2}\sigma^2 p_\mu}{p^4}.
\end{eqnarray}
The mixed propagator of the ghost sector is the one between $\omega$ and $\overline{c}$, and not the one between $\overline{\omega}$ and $c$, which vanishes: the mixing term of the bilinear action couples $\overline{\omega}$ to $c$, so that $\omega$ reaches $\overline{c}$ by propagating first into $\overline{\omega}$ and then, through this vertex, into $c$, while the opposite ordering would require the propagators $\langle \overline{\omega}\,\overline{\omega}\rangle$ and $\langle cc\rangle$, both of which are absent.

The ghost propagator remains of the simple form $1/p^2$, which is a direct consequence of the BRST symmetry and of the Landau gauge condition, the mixing term not feeding back into it. Importantly, the ghost does not acquire an anomalous dimension beyond the usual one, as ensured by the algebraic constraints encoded in the ghost equation, which provides a nontrivial check of the consistency of the BRST-preserving Gribov--Zwanziger construction. The remaining mixed propagators vanish identically, namely $\langle A\omega\rangle$, $\langle A\overline{\omega}\rangle$, $\langle\overline{\omega}c\rangle$ and all the mixed propagators between the scalar and the vector sectors, the corresponding terms of the action being trilinear in the fluctuations.

In the case of non-zero $a_{7}$, the term (\ref{refined}) contributes the mass $a_{7}\nu^{2}$ to the gauge field, and the propagators of the transverse sector change to
\begin{eqnarray}
\langle A_\mu^a(p) A_\nu^b(-p) \rangle &=& - \delta^{ab}\frac{p^2}{p^4 + a_7\nu^2 p^2 + \gamma^4} \, P_{\mu\nu}(p)\nonumber \\
\langle A_\mu^a(p) \varphi_\nu^b(-p) \rangle &=& -\delta^{ab} \, \frac{a_2 \sigma^{2}}{p^4 + a_7\nu^2 p^2 + \gamma^4} \, P_{\mu\nu}(p)\nonumber \\
\langle A_\mu^a(p) \overline{\varphi}_\nu^b(-p) \rangle &=& -\delta^{ab} \, \frac{a_1 \sigma^{2}}{p^4 + a_7\nu^2 p^2 + \gamma^4} \, P_{\mu\nu}(p),\nonumber \\
\langle \varphi_\mu^a(p) \overline{\varphi}_\nu^b(-p) \rangle &=& \delta^{ab} \, \frac{p^{4}+a_{7}\nu^{2}p^{2}+\frac{1}{2}\gamma^{4}}{p^{2}\bigl(p^4 + a_7\nu^2 p^2 + \gamma^4\bigr)}P_{\mu\nu}(p) + \delta^{ab}\frac{1}{p^{2}}L_{\mu\nu}(p)
\end{eqnarray}
The term (\ref{refined}) therefore supplies, through the same condensate and in a manner compatible with the BRST symmetry and with the ghost equation, a mass for the gauge field of the kind employed in the refined scenario. It should be observed, on the other hand, that the resulting propagator still vanishes at $p^{2}=0$, the decoupling behaviour of \cite{Dudal:2008sp}, in which the propagator attains a non-vanishing value at the origin, requiring in addition a condensate of the auxiliary fields of the type $\overline{\varphi}\varphi$, whose construction within the present formulation we leave for a future investigation.

\subsection{Decomposition into complex poles}

For spectral analysis purposes, the propagators involving $p^4 + \gamma^4$ can be decomposed into simple poles at $p^2 = \pm i\gamma^2$, as follows:

\begin{eqnarray}
\frac{p^2}{p^4 + \gamma^4} = \frac{1}{2} \left( \frac{1}{p^2 + i\gamma^2} + \frac{1}{p^2 - i\gamma^2} \right),
\end{eqnarray}

\begin{eqnarray}
\frac{1}{p^4 + \gamma^4} = \frac{1}{2i\gamma^2} \left( \frac{1}{p^2 + i\gamma^2} - \frac{1}{p^2 - i\gamma^2} \right),
\end{eqnarray}

This decomposition reveals the presence of complex conjugate poles in the propagator, a hallmark of the Gribov--Zwanziger confinement scenario. These complex poles are not associated with physical particle states, but rather indicate the absence of asymptotic gauge states in the physical spectrum. The BRST-invariant formulation presented here preserves this infrared feature while maintaining a consistent quantum interpretation.

\section{Renormalization}

\subsection{Functional identities and the quantum action principle}

As is easy to see, the kinetic term of the vector auxiliary fields is not symmetric, carrying one ordinary derivative and one covariant derivative, and it is therefore necessary to control the form of this operator, ensuring that it is maintained under perturbative corrections. The direct way to do this is to obtain an equation compatible with the quantum action principle \cite{Piguet:1995er,Henneaux:1994lbw,Becchi:1975nq} that guarantees the particular form of the operator. To this end, we introduce an insertion $\phi A^{a}_{\mu}$. As is standard for the study of an action in the presence of an insertion, we need a doublet of sources and equations to control it, and the standard way to introduce the insertion is
\begin{eqnarray}
s\{K^{a}_{\mu}\phi A^{a}_{\mu}\}&=&N^{a}_{\mu}\phi A^{a}_{\mu} - K^{a}_{\mu}\psi A^{a}_{\mu}+ K^{a}_{\mu}\phi (D_{\mu}c)^{a}\nonumber \\
s K^{a}_{\mu}&=&N^{a}_{\mu}; \,\,\, s N^{a}_{\mu}=0.
\end{eqnarray}
With this insertion term in the original action, we directly obtain the equations
\begin{eqnarray}
\int d^{4}x\{ \frac{\delta S}{\delta \overline\varphi^{a}_{\mu}} - a_{2}gm\frac{\delta S}{\delta N^{a}_{\mu}} \} &=&0\nonumber \\
\int d^{4}x \{ \frac{\delta S}{\delta c^{a} } -igf^{abc} \overline{c}^{b}\frac{\delta S}{\delta b^{c} } + gf^{abc} K^{b}_{\mu}\frac{\delta S}{\delta N^{c}_{\mu}} \} &=&0 \nonumber \\
\int d^{4}x\{ \frac{\delta S}{\delta \overline\omega^{a}_{\mu}}-gf^{abc} c^{b}\frac{\delta S}{\delta \overline{\varphi}^{c}_{\mu}}-a_{2}gm\frac{\delta S}{\delta K^{a}_{\mu}}\}&=&0.
\end{eqnarray}
These equations are compatible with the quantum action principle, and serve not only to control the insertion but also to guarantee the form of the kinetic operator as desired. It is easy to see that, although the ghost equation itself changes when the insertion sources are set to zero, the identity is trivially maintained; furthermore, the property of ensuring a renormalization fixed point for the ghost, obtained from the original ghost equation, is preserved as well, as we will see from the ghost renormalization result obtained later on.

The renormalizability of the model is established via the algebraic renormalization method, which relies on a set of functional identities: the Slavnov--Taylor identity, the ghost equation, the antighost equation, the $U(1)$ Ward identity, and the $X$-symmetry. These identities constrain the possible counterterms and guarantee that the action remains stable under quantum corrections.

The source term takes the form
\begin{eqnarray}
S_{j}&=& \int d^{4}x \{ -\Omega^{a\mu}(D_{\mu}c)^{a} + L^{a}\frac{g}{2}f^{abc}c^{b}c^{c} \nonumber \\
&+& J^{a\mu} gf^{abc}c^{b}\overline{\varphi}_{\mu }^{c} + \overline{J}^{a\mu}(\omega _{\mu }^{a}+gf^{abc}c^{b}\varphi_{\mu }^{c}) \nonumber \\
&+&\chi^{a\mu}( \overline{\varphi}_{\mu }^{a} + gf^{abc}c^{b}\overline{\omega}_{\mu }^{c}) +\overline{\chi}^{a\mu}gf^{abc}c^{b}\omega^{c}_{\mu})\nonumber\\
&+& N^{a}_{\mu}\phi A^{a}_{\mu} - K^{a}_{\mu}\psi A^{a}_{\mu}+ K^{a}_{\mu}\phi (D_{\mu}c)^{a} \}
\end{eqnarray}
and we can now extend the action, and the functional symmetries along with it, to take the sources into account, defining the fully quantized action $\Sigma=S+S_{j}$:

\begin{eqnarray}
\Sigma &=& S + S_{j}
\end{eqnarray}
where it is understood that $S_{j}$ includes the insertion already discussed at the beginning of this section.

The $U(1)$ symmetry, previously presented in global form, is written locally through the operator $U$ acting on $\Sigma$:
\begin{eqnarray}
U(\Sigma)&=& \int d^{4}x \{ \overline{\varphi}^{a}_{\mu}\frac{\delta \Sigma}{\delta \overline{\varphi}^{a}_{\mu}}+ \overline{\omega}^{a}_{\mu}\frac{\delta \Sigma}{\delta \overline{\omega}^{a}_{\mu}} + \overline{\psi}\frac{\delta \Sigma}{\delta \overline{\psi}}+ \overline{\phi}\frac{\delta \Sigma}{\delta \overline{\phi}}\nonumber \\
&-&\varphi^{a}_{\mu}\frac{\delta \Sigma}{\delta \varphi^{a}_{\mu}} -\omega^{a}_{\mu}\frac{\delta \Sigma}{\delta \omega^{a}_{\mu}}- \psi\frac{\delta \Sigma}{\delta \psi} - \phi\frac{\delta \Sigma}{\delta \phi}\nonumber \\
&+&J^{a\mu}\frac{\delta \Sigma}{\delta J^{a\mu}} + \chi^{a\mu}\frac{\delta \Sigma}{\delta \chi^{a\mu}} - \overline{J}^{a\mu}\frac{\delta \Sigma}{\delta \overline{J}^{a\mu}} - \overline{\chi}^{a\mu}\frac{\delta \Sigma}{\delta \overline{\chi}^{a\mu}} + K^{a}_{\mu}\frac{\delta \Sigma}{\delta K^{a}_{\mu}} + N^{a}_{\mu}\frac{\delta \Sigma}{\delta N^{a}_{\mu}}\} = 0
\end{eqnarray}

We also have another fundamental symmetry, $X$, an operator carrying ghost number $-1$ that, given its ghost number and functional form, must be the generator of the $U$ symmetry.

\begin{eqnarray}
X(\Sigma)&=&\int d^{4}x \{ \varphi^{a}_{\mu}\frac{\delta \Sigma}{\delta \omega^{a}_{\mu}}-\overline{\omega}^{a}_{\mu}\frac{\delta \Sigma}{\delta \overline{\varphi}^{a}_{\mu}} - \overline{\psi}\frac{\delta \Sigma}{\delta \overline{\phi}}+\phi\frac{\delta \Sigma}{\delta \psi}\nonumber \\
&-& \overline{\chi}^{a\mu} \frac{\delta \Sigma}{\delta \overline{J}^{a\mu}} -J^{a\mu}\frac{\delta \Sigma}{\delta \chi^{a\mu}} - K^{a}_{\mu}\frac{\delta \Sigma}{\delta N^{a}_{\mu}} \} = \int d^{4}x \{ J^{a\mu}\overline{\varphi}^{a}_{\mu} + \overline{\chi}^{a\mu}\omega^{a}_{\mu} \}\label{XS2}
\end{eqnarray}

It is also possible to write down the symmetry equations for the vector auxiliaries and for the ghost $c^{a}$; the ghost equation, incidentally, now carries a functional source term, but it is exactly the same ghost equation as in the Landau case without sources, since the sources can be set to zero for quantum calculations, and it is clear that this is the same ghost equation that holds in the Landau gauge.

\begin{eqnarray}
F^{a}_{\mu}(\Sigma)&=&\int d^{4}x\{ \frac{\delta \Sigma}{\delta \overline\varphi^{a}_{\mu}} - a_{2}gm\frac{\delta \Sigma}{\delta N^{a}_{\mu}} \} = \int d^{4}x\{ gf^{abc}c^{b}J^{c}_{\mu}\}
\end{eqnarray}

\begin{eqnarray}
G^{a}(\Sigma)&=& \int d^{4}x \{ \frac{\delta \Sigma}{\delta c^{a} } -igf^{abc} \overline{c}^{b}\frac{\delta \Sigma}{\delta b^{c} } + gf^{abc} K^{b}_{\mu}\frac{\delta \Sigma}{\delta N^{c}_{\mu}}\}\nonumber \\
&=&\int d^{4}x \{ gf^{abc} \{- J^{b\mu}\overline{\varphi}^{c}_{\mu} - \overline{J}^{b\mu}\varphi^{c}_{\mu}- L^{b}c^{c} + \chi^{b\mu}\overline{\omega}^{c}_{\mu}+ \overline{\chi}^{b\mu}\omega^{c}_{\mu} + \Omega^{b}_{\mu}A^{c\mu}\}\}
\end{eqnarray}

\begin{eqnarray}
H^{a}_{\mu}(\Sigma)&=& \int d^{4}x\{ \frac{\delta \Sigma}{\delta \overline\omega^{a}_{\mu}}-gf^{abc} c^{b}\frac{\delta \Sigma}{\delta \overline{\varphi}^{c}_{\mu}}-a_{2}gm\frac{\delta \Sigma}{\delta K^{a}_{\mu}}- g f^{abc}J^{b}_{\mu}\frac{\delta \Sigma}{\delta L^{c}}\}\nonumber \\
&=& \int d^{4}x\{- gf^{abc}\chi^{b}_{\mu}c^{c}\}
\end{eqnarray}

Given that all fields and sources carrying group indices can be collected into a total element $\theta^{a}= (\varphi^{a\mu}, \overline{\varphi}^{a\mu}, \omega^{a\mu}, ...)$, we have
\begin{equation}
W^{a}(\Sigma)=\int d^{4}x \{gf^{abc}\theta^{b}\frac{\delta \Sigma}{\delta \theta^{c}}\} = 0 .
\end{equation}

\begin{eqnarray}
\frac{\delta \Sigma}{\delta b^{a}}&=&i\partial_{\mu}A^{a\mu}\nonumber \\
\overline{G}^{a}(\Sigma)&=&\partial_{\mu}\frac{\delta \Sigma}{\delta \Omega^{a}_{\mu}} + \frac{\delta \Sigma}{\delta \overline{c}^{a}} = 0
\end{eqnarray}

The Slavnov--Taylor identity then reads
\begin{eqnarray}
S(\Sigma) &=& \int d^{4}x \{ \frac{\delta \Sigma}{\delta A^{a}_{\mu}}\frac{\delta \Sigma}{\delta \Omega^{a}_{\mu}} +
\frac{\delta \Sigma}{\delta c^{a}}\frac{\delta \Sigma}{\delta L^{a}} + \frac{\delta \Sigma}{\delta \overline{\varphi}^{a}_{\mu}}\frac{\delta \Sigma}{\delta J^{a}_{\mu}}+ \frac{\delta \Sigma}{\delta \varphi^{a}_{\mu}}\frac{\delta \Sigma}{\delta \overline{J}^{a}_{\mu}}+ \frac{\delta \Sigma}{\delta \overline{\omega}^{a}_{\mu}}\frac{\delta \Sigma}{\delta \chi^{a}_{\mu}} +
\frac{\delta \Sigma}{\delta \omega^{a}_{\mu}}\frac{\delta \Sigma}{\delta \overline{\chi}^{a}_{\mu}} \nonumber \\
&+& \overline{\phi}\frac{\delta \Sigma}{\delta \overline{\psi}} + \psi\frac{\delta \Sigma}{\delta \overline{\phi}} + N^{a}_{\mu}\frac{\delta \Sigma}{\delta K^{a}_{\mu}}+ ib^{a}\frac{\delta \Sigma}{\delta\overline{c}^{a}}\} = 0 .
\end{eqnarray}

and, in linearized form,\footnote{We denote by $\beta$ the linearized Slavnov-Taylor operator, which extends the BRST symmetry to the external sources; on the fields, $\beta$ reduces to the classical BRST operator $s$.}
\begin{eqnarray}
\beta&=& \int d^{4}x \{ \frac{\delta \Sigma}{\delta A^{a}_{\mu}}\frac{\delta }{\delta \Omega^{a}_{\mu}} + \frac{\delta \Sigma}{\delta \Omega^{a}_{\mu}}\frac{\delta }{\delta A^{a}_{\mu}} + \frac{\delta \Sigma}{\delta c^{a}}\frac{\delta }{\delta L^{a}}
+ \frac{\delta \Sigma}{\delta L^{a}}\frac{\delta }{\delta c^{a}}\nonumber \\
&+& \frac{\delta \Sigma}{\delta \overline{\varphi}^{a}_{\mu}}\frac{\delta }{\delta J^{a}_{\mu}} +
\frac{\delta \Sigma}{\delta J^{a}_{\mu}}\frac{\delta}{\delta \overline{\varphi}^{a}_{\mu}} +
\frac{\delta \Sigma}{\delta \varphi^{a}_{\mu}}\frac{\delta }{\delta \overline{J}^{a}_{\mu}} +
\frac{\delta \Sigma}{\delta \overline{J}^{a}_{\mu}}\frac{\delta }{\delta \varphi^{a}_{\mu}}+
\frac{\delta \Sigma}{\delta \overline{\omega}^{a}_{\mu}}\frac{\delta }{\delta \chi^{a}_{\mu}} +
\frac{\delta \Sigma}{\delta \chi^{a}_{\mu}} \frac{\delta }{\delta \overline{\omega}^{a}_{\mu}}\nonumber \\
&+& \frac{\delta \Sigma}{\delta \omega^{a}_{\mu}}\frac{\delta }{\delta \overline{\chi}^{a}_{\mu}} +
\frac{\delta \Sigma}{\delta \overline{\chi}^{a}_{\mu}} \frac{\delta }{\delta \omega^{a}_{\mu}} +
\overline{\phi}\frac{\delta }{\delta \overline{\psi}} + \psi\frac{\delta }{\delta \overline{\phi}} + N^{a}_{\mu}\frac{\delta }{\delta K^{a}_{\mu}}
+ ib^{a}\frac{\delta }{\delta\overline{c}^{a}}\}
\end{eqnarray}
The relevant commutator relations then read
\begin{eqnarray}
\{\beta,X\}&=& -U\nonumber \\
\{ G^{a},\beta\}&=& W^{a}\nonumber \\
\left[\frac{\delta}{\delta b^{a}},\beta \right]&=& i \overline{G}^{a}\nonumber \\
\left[ F^{a}_{\mu},\beta \right] &=& H^{a}_{\mu}
\end{eqnarray}
The relation $\{\beta, X\} = -U$ is of central importance for the renormalization of the model. It implies that the $U(1)$ symmetry is not an independent requirement: once the BRST symmetry is preserved at the quantum level and the $X$-symmetry is compatible with renormalization, the $U(1)$ Ward identity follows automatically. Consequently, the renormalization of the model does not require the introduction of additional counterterms to enforce $U(1)$ invariance, which greatly simplifies the algebraic renormalization procedure.

\subsection{The counterterm and the renormalization constants}
\label{sec:counterterm}
Before writing down the most general form of the counterterm action, it is appropriate to present a table of fields and sources. The row $\DimUV$ refers to the ultraviolet dimension of each field, $N_{ghost}$ to its ghost number, and $U(1)$ to the charge under the $U(1)$ symmetry already presented earlier.

\begin{table}[htbp]
\centering
\caption{Fields}
\label{tab:12colunas}
\newcolumntype{C}{>{\centering\arraybackslash$}p{0.6cm}<{$}}
\begin{tabular}{c*{12}{C}}
\toprule
& A^{a}_{\mu} & \overline{c}^{a} & c^{a} & b^{a} & \overline{\omega}^{a}_{\mu} & \overline{\varphi}^{a}_{\mu} & \omega^{a}_{\mu} & \varphi^{a}_{\mu} & \overline{\psi} & \overline{\phi} & \psi & \phi \\
\midrule
$\DimUV$ & 1 & 2 & 0 & 2 & 1 & 1 & 1 & 1 & 1 & 1 & 1 & 1 \\
$N_{ghost}$ & 0 & -1 & 1 & 0 & -1 & 0 & 1 & 0 & -1 & 0 & 1 & 0 \\
$U(1)$ & 0 & 0 & 0 & 0 & 1 & 1 & -1 & -1 & 1 & 1 & -1 & -1 \\
\bottomrule
\end{tabular}

\centering
\caption{Sources}
\label{tab:8colunas}
\newcolumntype{D}{>{\centering\arraybackslash$}p{0.6cm}<{$}}
\begin{tabular}{c*{8}{D}}
\toprule
&\Omega^{a}_{\mu} & L^{a} & J^{a}_{\mu} & \chi^{a}_{\mu} & \overline{J}^{a}_{\mu} & \overline{\chi}^{a}_{\mu} & K^{a}_{\mu} & N^{a}_{\mu} \\
\midrule
$\DimUV$ & 3 & 4 & 3 & 3 & 3 & 3 & 2 & 2 \\
$N_{ghost}$ & -1 & -2 & -1 & 0 & -1 & -2 & -1 & 0 \\
$U(1)$ & 0 & 0 & 1 & 1 & -1 & -1 & 1 & 1 \\
\bottomrule
\end{tabular}
\end{table}

The most general counterterm compatible with the functional identities of the preceding subsection, of ultraviolet dimension four and vanishing ghost number, takes the form $\Gamma_{c}=\Gamma^{0}+\beta\Gamma^{-1}$, where $\Gamma^{0}$ is the Yang--Mills term, with coefficient $x_{0}$, and $\Gamma^{-1}$ is an integrated local polynomial of ghost number $-1$ whose coefficients $x_{i}$ are constrained only by the consistency equations imposed on $\Gamma_{c}$ itself. The explicit form of $\Gamma_{c}$, the redefinitions of the fields, sources and parameters, and the complete solution of the resulting system for the renormalization constants are collected in Appendix~\ref{app:renorm}; here we retain only the results that bear on the interpretation of the model. We write the renormalization factors as $Z_\Phi=1+\epsilon z_{\Phi}+O(\epsilon^{2})$, where $\Phi$ denotes any field, parameter, or source of the theory and $\epsilon$ is the expansion parameter.

Two entries of the solution deserve to be singled out, because the order in which they are read matters. The monomial $m^{2}(\overline{\phi}\phi-\overline{\psi}\psi)$ of the classical action is matched by the coefficient $x_{8}$, which occurs nowhere else in $\beta\Gamma^{-1}$, Eq.~(\ref{eq:betaGamma}), so that this monomial alone determines the renormalization of the mass,
\begin{equation}
z_{m}=\frac{x_{8}-x_{7}}{2}.
\label{eq:zm}
\end{equation}
The two monomials carrying $a_{1}gm$ and $a_{2}gm$, once the sum rules $z_{\overline{\phi}}+z_{\phi}=x_{7}$ and $z_{\varphi}+z_{\overline{\varphi}}=x_{5}+x_{12}$ are used, then give
\begin{equation}
z_{1}+z_{2}+2z_{g}+2z_{m}+x_{7}+x_{5}+x_{12}=\tilde{x}_{16}+x_{10}-x_{0},
\end{equation}
in which the two factors $z_{g}=-x_{0}/2$ are those supplied by the explicit powers of the coupling in these terms, and which, with (\ref{eq:zm}), determines
\begin{equation}
z_{1}+z_{2}= \tilde{x}_{16}+x_{10}-x_{8}-x_{5}-x_{12}.
\label{eq:z1z2}
\end{equation}

It should be observed that $z_{1}$ and $z_{2}$ are not separately meaningful, each of them depending on the normalization of the auxiliary pair discussed in Appendix~\ref{app:renorm} while their sum does not, which is the algebraic counterpart of the fact that only the product $a_{1}a_{2}$ enters the propagators of Section 4 and the scale $\gamma^{4}$.

The relation $z_7 = -x_0 - x_7$, obtained in Appendix~\ref{app:renorm}, shows that the coupling $a_7$, which controls the refinement term, is not a renormalization-group invariant; its renormalization factor is given by $Z_7 = 1 - \epsilon (x_0 + x_7)$, implying that $a_7$ is an essential parameter of the theory, whose value at a given scale must be fixed by a physical condition (e.g., the infrared gauge mass gap). This is precisely the same status enjoyed by the refinement parameter in the literature \cite{Dudal:2008sp}. The coefficients $a_1$ and $a_2$, on the other hand, are not separately meaningful, only the renormalization of their product being independent of the normalization of the auxiliary pair, and whether it vanishes or follows that of $g^{2}$ is decided by (\ref{eq:z1z2}), that is, by the coefficients of the divergences themselves.

The remaining $z$-factors are expressed in terms of the independent parameters $x_0, x_1, x_5, x_7, x_{10}, x_{12},$ and $\tilde{x}_{16}$, as given in Appendix~\ref{app:renorm}. These relations demonstrate the algebraic consistency of the renormalization program and confirm that the model is renormalizable without introducing new independent counterterms beyond those already present in the classical action.

The relation (\ref{eq:z1z2}) settles a question of interpretation which is worth stating carefully. Whether the coefficients $a_{1}$ and $a_{2}$ may be held fixed as pure numbers, which corresponds to $z_{1}+z_{2}=0$ and hence, by (\ref{eq:z1z2}), to $x_{8}=\tilde{x}_{16}+x_{10}-x_{5}-x_{12}$, is not a matter of choice but a condition on the coefficients of the divergences, and deciding whether it holds requires the explicit computation of $\tilde{x}_{16}$, $x_{10}$, $x_{8}$, $x_{5}$ and $x_{12}$ at one loop, which we have not carried out here. Had the coupling not been written explicitly in the mixed terms of the action, the same condition would read $x_{8}=\tilde{x}_{16}+x_{10}-x_{0}-x_{5}-x_{12}$, the two differing by $2z_{g}=-x_{0}$, which is the algebraic trace of the power of $g$ carried by those terms.

It should be stressed that this bears on the parameters alone and not on the cohomology, since $a_{1}$ and $a_{2}$ are coefficients of monomials of the action and appear nowhere in the BRST transformations: the doublet structure of the quartets, and with it the classification of the observables discussed in Section~\ref{sec:cohomology}, is insensitive to their values and to the way they are written.

% ============================================================================
% BROKEN PHASE (former Appendix A)
% ============================================================================

\section{The broken phase: Ward identities, BRST symmetry and vacuum condition}
\label{sec:brokenWard}

In this section we clarify how the functional identities used in the algebraic
renormalization are to be read once the scalar fields acquire a vacuum
expectation value. The main point is that no modification of the operator
algebra is required, since the identities are formulated for the complete action
$\Sigma$ before any vacuum is chosen. Spontaneous symmetry breaking only means
that, when the fields are shifted by their expectation values, the Ward
identities acquire constant one-point contributions, and the minimization of the
effective potential is precisely the condition that these contributions vanish
on shell.

\subsection{Shifted fields and parameters}

We write
\begin{equation}
\phi=\nu+\eta,\qquad \bar\phi=\nu+\bar\eta,\qquad
\langle \eta\rangle=\langle\bar\eta\rangle=0,
\label{eq:app-shift}
\end{equation}
with $\psi$ and $\bar\psi$ having vanishing expectation values. As in Section 4,
\begin{equation}
\nu^2=\frac{m^2}{\lambda},\qquad \sigma^2=gm\nu,\qquad
\gamma^4=2a_1a_2g^2m^2\nu^2 .
\label{eq:app-scales}
\end{equation}
The action obtained by replacing $(\phi,\bar\phi)$ by $(\nu+\eta,\nu+\bar\eta)$
will be denoted by $\Sigma_\nu[\eta,\bar\eta,\dots]$, where the ellipsis stands
for all remaining fields and sources. The shift does not change the functional
algebra of the operators; it only reorganizes the arguments of the generating
functional.

\subsection{The operator algebra is not shifted}

The BRST operator, the $X$-symmetry and the $U(1)$ Ward operator are defined at
the level of the complete action $\Sigma$, with the external sources included.
Their algebraic relation remains
\begin{equation}
\{\beta,X\}=-U .
\label{eq:app-algebra}
\end{equation}
In particular, we do not redefine $U$ or $X$ by terms proportional to $\nu$.
Such redefinitions are neither needed nor consistent at the operator level,
because they mix Ward operators with vacuum-dependent constants.

What changes in the broken phase is not the algebraic relation (\ref{eq:app-algebra}) itself, but the
fact that the vacuum is not annihilated by the $U(1)$ charge; operationally, this
appears after the shift as constant terms in the one-point Ward identities.

\subsection{The BRST symmetry in the non-trivial vacuum}
\label{sec:cohomology}

We now come to the point which is the conceptual core of the construction, and which
deserves to be stated with some care, since a superficial reading of the condensation
of a quartet suggests that the BRST symmetry has been lost. It has not. The symmetry
is not broken when the scalar fields acquire their vacuum expectation values: what
the displacement of the vacuum does is to select the realization in which the
symmetry is manifest, the generator appropriate to the broken phase being the vacuum
operator $s_{\nu}$ obtained by substituting the displaced fields
(\ref{eq:app-shift}) into the original transformations (\ref{quartet}),
\begin{eqnarray}
\label{eq:app-snu}
s_{\nu}\overline{\psi} &=& \nu + \overline{\eta}, \qquad s_{\nu}\overline{\eta} = 0,
\nonumber \\
s_{\nu}\eta &=& \psi, \qquad\qquad\;\, s_{\nu}\psi = 0,
\end{eqnarray}
the transformations of all the remaining fields being left untouched by the
displacement. This operator is nilpotent and is an exact symmetry of $\Sigma_{\nu}$,
in the sense of the hidden symmetries of Coleman, and it is with respect to it, or to
the operator obtained from it by a filtration, that the observables of the broken
phase are to be classified. We have developed this mechanism in detail in
\cite{Amaral:2022,Amaral:2023}, where the transition operator was constructed for
spontaneously broken gauge theories and the cohomology of the broken phase was
explicitly obtained; in fact, the cohomology of a nilpotent operator is contained in
the cohomology of any nilpotent operator obtained from it by a filtration, so that
the observables of a given phase are to be found among those of the preceding one.

Two statements must then be carefully distinguished, and it is their confusion that
gives the impression of a broken symmetry. The first concerns the cohomology of local
functionals, and here nothing at all is lost. The displacement
(\ref{eq:app-shift}) is an invertible change of variables, polynomial in both
directions, and since $\nu$ carries ultraviolet dimension one and enters only with
non-negative powers, it maps the space of integrated local polynomials of dimension
four and definite ghost number onto itself. As $s_{\nu}$ is by construction the image
of $s$ under this map, the two complexes are isomorphic and
\begin{equation}
H(s_{\nu}) = H(s).
\end{equation}
The pairs $(\overline{\psi},\overline{\eta})$ and $(\eta,\psi)$ remain doublets, as
is immediate from (\ref{eq:app-snu}), the scalar sector stays cohomologically
trivial, and the observables of the model are, in either phase, those of the
Yang--Mills cohomology.

It is instructive to see why the present situation is simpler in this respect than
that of a spontaneously broken gauge theory. There the field that condenses is a
matter field whose BRST transformation is proportional to the Faddeev--Popov ghost,
so that the vacuum operator acquires an inhomogeneous term of the form
$s_{\nu}\chi^{a}=\nu gf^{abz}c^{b}+\ldots$, which no redefinition can remove: the
would-be Goldstone modes and the ghosts of the broken directions are paired into
doublets that did not exist in the symmetric phase, and the cohomology of the broken
phase is genuinely different from the symmetric one \cite{Amaral:2022,Amaral:2023}.
In the model considered here, on the other hand, the scalar quartet is a colour
singlet, and its transformations (\ref{quartet}) are the only ones of the whole set
of fields in which the ghost does not appear. This is the same feature that
guarantees the ghost equation of Section 5, and it implies that no field whose
transformation involves the ghost acquires a vacuum expectation value, so that no
such re-pairing takes place. Put in another way, the displacement of the vacuum does
not create new observables here; it fixes the value of the scale that appears in the
two-point functions of Section 4.

The second statement concerns the vacuum itself, and it is here that the choice of
phase makes itself felt. Since $\langle\overline{\eta}\rangle=0$ while
$\langle\overline{\phi}\rangle=\nu$, the global $U(1)$ charge does not annihilate the
vacuum, as we have already observed, and the same is true of the operator obtained by
projecting the one-point identities onto the vacuum. One should note that this is
precisely the structure met in the standard Higgs phase, where the rigid gauge Ward
operator satisfies $\{G^{a},\beta\}=W^{a}$ and is spontaneously broken by the
condensate without any breaking of the BRST symmetry being implied. In the present
case the relation is $\{\beta,X\}=-U$, and the situation is the same: $X$ and $G^{a}$
are Ward operators of the functional type, not local charges acting on the state
space, and the fact that a broken charge can be written as an anticommutator with
$\beta$ does not by itself make the BRST symmetry broken.

What remains, then, is the characterization of the modes, and here a feature of the
present construction deserves to be pointed out, since it distinguishes it from the
usual realization of the quartet mechanism. Because $s_{\nu}\overline{\psi}$ contains
the constant $\nu$, the operator (\ref{eq:app-snu}) does not preserve the counting of
fields, and the quadratic part of the action is not by itself invariant. Indeed, with
$\phi=\nu+\eta$ and $\overline{\phi}=\nu+\overline{\eta}$, the quadratic part of the
potential is $\frac{m^{2}}{2}(\eta+\overline{\eta})^{2}$ and its variation,
$m^{2}(\eta+\overline{\eta})\psi$, is cancelled not by another quadratic term but by
the variation of the cubic vertex
$\lambda\nu(\eta+\overline{\eta})(\overline{\eta}\eta-\overline{\psi}\psi)$, through
the term $\nu\psi$ contained in $s_{\nu}(\overline{\psi}\psi)$. The invariance
therefore closes only between orders, and no linearized operator is available for the
free theory. This is why the mass eigenmodes of Section 4 are not degenerate with
their partners under (\ref{eq:app-snu}): the massive combination
$\eta+\overline{\eta}$ has $2m^{2}$, while the fields into which it is mapped are
massless.

The consequence is that the removal of the massless modes from the physical subspace
cannot be argued here by the free-field counting of quartets, as it can in the
electroweak case, where the vev-induced term of the transition operator is linear in
the fields and the quadratic action is separately invariant. What the algebraic
analysis does establish is the statement of the preceding paragraphs: the symmetry is
exact in both phases and the cohomology through which the observables are defined is
the same, the non-trivial vacuum providing only the value of $\nu$, fixed by the
minimization of $V_{\rm eff}$. The characterization of the asymptotic states in the
interacting theory, and in particular the fate of the massless modes of the scalar
sector, requires the analysis of the full charge and is left for a separate
investigation.

\subsection{Shifted Ward identities and tadpoles}

Acting with $U$ on the shifted action gives
\begin{equation}
U(\Sigma_\nu)
=
\left.U(\Sigma)\right|_{\phi=\nu+\eta,\bar\phi=\nu+\bar\eta}
=
U_{\rm fl}(\Sigma_\nu)
+\nu\int d^4x\left(
\frac{\delta\Sigma_\nu}{\delta\bar\eta}
-\frac{\delta\Sigma_\nu}{\delta\eta}
\right),
\label{eq:app-Ushift}
\end{equation}
where $U_{\rm fl}$ is the same charge operator restricted to the fluctuating
fields. Since the original identity is $U(\Sigma)=0$, (\ref{eq:app-Ushift})
implies
\begin{equation}
U_{\rm fl}(\Sigma_\nu)
=
-\nu\int d^4x\left(
\frac{\delta\Sigma_\nu}{\delta\bar\eta}
-\frac{\delta\Sigma_\nu}{\delta\eta}
\right).
\label{eq:app-Ufl}
\end{equation}
The right-hand side is a tadpole-type contribution. At the physical vacuum it
must vanish by the minimization condition,
\begin{equation}
\left.
\frac{\delta\Sigma_\nu}{\delta\eta}
\right|_{\rm sources=0}
=
\left.
\frac{\delta\Sigma_\nu}{\delta\bar\eta}
\right|_{\rm sources=0}
=0,
\end{equation}
or, at the quantum level,
\begin{equation}
\left.\frac{\partial V_{\rm eff}}{\partial\nu}\right|_{\nu=\nu_0}=0 .
\label{eq:app-min}
\end{equation}
Thus the broken-phase Ward identities are compatible with the same operator
relation (\ref{eq:app-algebra}); the constant pieces generated by the shift are
removed by the vacuum condition rather than by a redefinition of the symmetry
generators.

An analogous statement holds for $X$. Because $X$ maps $\phi$ into $\psi$ and
$\psi$ into $\nu$-dependent constants after the shift, its action on
$\Sigma_\nu$ also produces one-point terms. These terms are not independent
data: the algebra (\ref{eq:app-algebra}) relates them to the $U(1)$ tadpoles above,
so that, once the vacuum condition (\ref{eq:app-min}) is imposed, the
$X$-identity reduces to the same source-dependent Ward identity used in the
renormalization analysis.

\subsection{The effective potential and the vacuum condition}

At the bilinear level the shifted action contains the combinations
\begin{equation}
a_1\sigma^2 A_\mu^a\varphi^{a\mu}
+a_2\sigma^2 A_\mu^a\bar\varphi^{a\mu},
\end{equation}
which reproduce the Gribov-type mixing and yield
\begin{equation}
\langle A_\mu^a(p)A_\nu^b(-p)\rangle
=
-\delta^{ab}\frac{p^2}{p^4+\gamma^4}P_{\mu\nu}(p),
\end{equation}
with $\gamma^4$ given by (\ref{eq:app-scales}). The vacuum condition
(\ref{eq:app-min}) determines $\nu$ dynamically, and it is instructive to work it out
at one loop, since it is here that the present mechanism differs most clearly from
the standard construction. At this order the dependence of the vacuum energy on $\nu$
comes from the quadratic operators evaluated at the constant background
$\phi=\bar\phi=\nu$, and it is enough to collect the sectors that the background
actually reaches.

In the transverse sector the fields $(A^{a}_{\mu},\varphi^{a}_{\mu},
\bar\varphi^{a}_{\mu})$ mix through the terms above, and the determinant of the
corresponding quadratic form is
\begin{equation}
-p^{2}\left(p^{4}+a_{7}\nu^{2}p^{2}+\gamma^{4}\right),
\end{equation}
so that the whole dependence on the vacuum enters through $\gamma^{4}$ and, for
non-vanishing $a_{7}$, through the gauge mass; the longitudinal components of
$\varphi$ and $\bar\varphi$ do not mix with the gauge field and are independent of
$\nu$. In the sector of $\omega$, $\bar\omega$ and the Faddeev--Popov ghosts, the
mixing term $a_{2}\sigma^{2}\bar\omega_{\mu}\partial^{\mu}c$ couples $\bar\omega$ to
$c$ alone, so that the quadratic form is triangular and its determinant does not
depend on $\nu$ at all: this sector drops out of the effective potential. In the
scalar sector, writing $\kappa=-m^{2}+\lambda\nu^{2}$ for the derivative of the
potential at the background, the modes of $\eta$ and $\bar\eta$ are governed by
$p^{2}+\kappa$ and $p^{2}+\kappa+2\lambda\nu^{2}$ while those of $\psi$ and
$\bar\psi$ are governed by $p^{2}+\kappa$, so that the cancellation between the
commuting and the anticommuting members of the quartet is complete only in the
symmetric vacuum. For $\nu\neq0$ the massive combination survives and the sector
contributes the usual Coleman--Weinberg term of a condensing scalar.

Collecting the three sectors, the one-loop effective potential reads, for $a_{7}=0$
and up to terms independent of the vacuum,
\begin{equation}
V_{\rm eff}(\nu)= -m^{2}\nu^{2}+\frac{\lambda}{2}\nu^{4}
+\frac{3(N^{2}-1)}{2}\int\frac{d^{4}q}{(2\pi)^{4}}\ln\left(q^{4}+\gamma^{4}\right)
+\frac{1}{2}\int\frac{d^{4}q}{(2\pi)^{4}}
\ln\frac{q^{2}+\kappa+2\lambda\nu^{2}}{q^{2}+\kappa},
\end{equation}
and the condition (\ref{eq:app-min}) becomes
\begin{equation}
\lambda\nu^{2}-m^{2}
+3(N^{2}-1)a_{1}a_{2}g^{2}m^{2}\int\frac{d^{4}q}{(2\pi)^{4}}\frac{1}{q^{4}+\gamma^{4}}
+\frac{\lambda}{2}\int\frac{d^{4}q}{(2\pi)^{4}}
\left[\frac{3}{q^{2}+\kappa+2\lambda\nu^{2}}-\frac{1}{q^{2}+\kappa}\right]=0 .
\label{eq:app-gap}
\end{equation}

Three features of (\ref{eq:app-gap}) delimit what the mechanism does and does not
accomplish. In the first place, the integral characteristic of the
Gribov--Zwanziger gap equation does appear, and with it the infrared structure of the
Gribov propagator, which confirms that the condensation of the quartet produces the
bilinear content of the horizon term. In the second place, its coefficient,
$3(N^{2}-1)a_{1}a_{2}g^{2}m^{2}$, is proportional to $g^{2}$, as is that of the
standard gap equation, this being the reason why the coupling was written explicitly
in the mixed terms of the action of Section 3: with $\gamma^{4}=2a_{1}a_{2}g^{2}m^{2}\nu^{2}$
the Gribov scale vanishes together with the coupling, as it must if it is to
originate in the restriction of the functional integral, whereas coefficients
independent of $g$ would leave the gauge propagator of Gribov type in a limit in
which no gauge copies are present. The counting of colour components, on the other
hand, is $N^{2}-1$ rather than the $N$ of the standard construction, which is the
direct consequence of the use of fields with a single adjoint index. Finally, and in
contrast with the standard construction, where the whole dependence of the vacuum
energy on the scale resides in the horizon term, the tree-level potential contributes
here the terms $\lambda\nu^{2}-m^{2}$, so that (\ref{eq:app-gap}) is not of the form
$\ldots=1$ but a Coleman--Weinberg condition in which the Gribov integral is one
contribution among others.

The scale is therefore determined by the vacuum of the scalar sector and by the
coupling jointly, through $\gamma^{4}=2a_{1}a_{2}g^{2}m^{2}\nu^{2}$ with $\nu$ given
by (\ref{eq:app-gap}), and not by the gauge dynamics alone as in the horizon
condition. We leave the detailed investigation of this condition for future work,
noting only that the Gribov integral is logarithmically divergent,
\begin{equation}
\int\frac{d^{4}q}{(2\pi)^{4}}\frac{1}{q^{4}+\gamma^{4}}
=\frac{1}{8\pi^{2}}\ln\frac{\Lambda}{\gamma},
\end{equation}
so that the vacuum condition requires renormalization, exactly as the horizon
condition does in the usual formulation.

\subsection{The nature of the scale}

Two properties of the vacuum condition deserve to be recorded separately, since they
concern the interpretation of the mechanism rather than its consistency.

The first is the direction in which the fluctuations of the gauge field act. Since
$\gamma^{4}$ is proportional to $\nu^{2}$, and the contribution of the transverse
sector to the effective potential increases with $\gamma^{4}$, that contribution
increases with the vacuum expectation value: for $a_{1}a_{2}>0$, which is what the
Gribov form of the propagator requires, the gauge fluctuations raise $V_{\rm eff}$ as
$\nu$ grows, and therefore act against the condensate. In (\ref{eq:app-gap}) this
appears as the positive sign of the gauge term, which displaces the solution towards
values of $\nu$ below the tree-level one and, for a sufficiently large coupling,
removes it altogether, the condensate and with it the Gribov scale being destroyed.
The argument just given establishes the sign of the effect and not its magnitude,
which depends on the renormalization of the logarithmically divergent integral; but
the sign alone is significant, for it is the reverse of what occurs in the standard
construction, where the horizon condition forces $\gamma$ to be different from zero
and the effect is governed by the growth of the coupling in the infrared. In this
sense the phase described here is one of weak coupling.

The second property concerns the origin of the scale itself. Using the tree-level
value $\nu^{2}=m^{2}/\lambda$ in $\gamma^{4}=2a_{1}a_{2}g^{2}m^{2}\nu^{2}$ gives
\begin{equation}
\gamma = \left(\frac{2a_{1}a_{2}}{\lambda}\right)^{1/4}\sqrt{g}\;m ,
\end{equation}
so that the Gribov scale is polynomial in the coupling and proportional to the mass
parameter of the scalar potential. This is to be contrasted with the standard
formulation, in which the horizon condition determines $\gamma$ in terms of the
coupling and the renormalization scale alone, through a mechanism of dimensional
transmutation. What the condensation of the quartet accomplishes, then, is to produce
the bilinear structure of the Gribov--Zwanziger action, and the propagator that goes
with it, with the scale inherited from the scalar sector rather than generated by the
gauge dynamics. Whether this is to be regarded as a limitation of the mechanism, or as
an indication that the scale of confinement should be sought in a hidden breaking of
the kind we have discussed in \cite{Amaral:2024}, is a question we prefer to leave
open.

\subsection{Remark on the ``central charge'' terminology}

If one projects the shifted one-point Ward identity onto the vacuum and uses the
gap equation, the constant term associated with the $U(1)$ variation is
proportional to $\gamma^2$. It can be informally viewed as a central-looking
constant in the broken-phase realization of the algebra. We stress, however,
that in the present formulation this is not an operatorial central extension of
$\{\beta,X\}$: the exact statement remains (\ref{eq:app-algebra}), and the
$\gamma^2$-dependent term is simply the vacuum value of a one-point Ward identity,
fixed by the minimization of $V_{\rm eff}$.

\subsection{Summary}

The broken-phase analysis can therefore be summarized as follows:
\begin{enumerate}
\item The functional identities are defined for $\Sigma$ before the choice of
vacuum, and the algebra $\{\beta,X\}=-U$ remains unchanged.
\item The symmetry is not lost in the broken phase. It is realized around the
displaced vacuum by the operator $s_{\nu}$ of (\ref{eq:app-snu}), which is nilpotent
and is still an exact symmetry of the action. Since the displacement is an invertible
polynomial change of variables, the cohomology of local functionals is the same in
both phases, $H(s_{\nu})=H(s)$, no re-pairing of the ghosts taking place because the
scalar quartet is the only sector whose transformations do not involve them.
\item Shifting $\phi$ and $\bar\phi$ by $\nu$ produces constant tadpole terms in
the $U(1)$ and $X$ Ward identities, not new symmetry generators.
\item The condition that these tadpoles vanish is the minimization of
$V_{\rm eff}$, which dynamically determines $\nu$.
\item Through $\sigma^2=gm\nu$ and $\gamma^4=2a_1a_2g^2m^2\nu^2$, this condition
reproduces the Gribov-type propagator, and at one loop it takes the form
(\ref{eq:app-gap}), in which the Gribov integral appears together with the
tree-level terms of the potential and the residual contribution of the massive
scalar, and not the form of the standard gap equation.
\item Any $\gamma^2$-dependent constant appearing after projection onto the
vacuum should be interpreted as the value of a one-point Ward identity, not as
an independent central extension of the BRST algebra.
\end{enumerate}

% ============================================================================
% SUFFICIENCY OF SSB
% ============================================================================

% ============================================================================
% SUFFICIENCY OF SSB
% ============================================================================

\section{Sufficiency of the Spontaneous Symmetry Breaking Mechanism for Reproducing the Gribov Condition}

In this section we elaborate on the argument that the spontaneous symmetry breaking (SSB) mechanism proposed above is sufficient to reproduce the Gribov condition, at least up to bilinear order, and that this sufficiency is, in fact, guaranteed by the structure of the Gribov-Zwanziger framework itself.

\subsection{The Gribov Condition and Its Local Realization}

The Gribov condition in the Landau gauge restricts the functional integral to the region where the Faddeev-Popov operator is positive definite:
\[
\mathcal{M}^{ab}[A] = -\partial_\mu D_\mu^{ab} > 0.
\]

In its original form, Gribov showed that this condition is equivalent to the no-pole condition on the ghost two-point function. For the color-singlet component of the ghost form factor, one has
\[
\sigma(0; A) = \frac{1}{d}\frac{N}{N^2-1}\int d^dx\, A_\mu^c(-\partial^2)^{-1}A_\mu^c < 1.
\]

The local realization of this condition, due to Zwanziger, introduces auxiliary fields and results in the horizon condition
\[
\langle H(A) \rangle = d(N^2-1),
\]
which fixes the Gribov parameter $\gamma$ through the gap equation
\[
\frac{3Ng^2}{4}\int \frac{d^4q}{(2\pi)^4}\frac{1}{q^4+\gamma^4} = 1.
\]

\subsection{The Bilinear Structure of the Gribov-Zwanziger Action}

The Gribov-Zwanziger action in its standard form contains auxiliary fields $\varphi_\mu^{ac}$ and $\bar\varphi_\mu^{ac}$ carrying two adjoint color indices. At bilinear order, the relevant terms are:
\[
S_{\text{GZ}}^{\text{bilinear}} = \int d^dx\left[ \bar\varphi_\mu^{ac}(-\partial^2)\varphi_\mu^{ac} + g\gamma^2 f^{abc}A_\mu^a(\varphi_\mu^{bc}+\bar\varphi_\mu^{bc}) - d(N^2-1)\gamma^4 \right].
\]

The color contraction $f^{abc}f^{dbc} = -N\delta^{ad}$ shows that the source term effectively couples the gluon to the antisymmetric combination of auxiliary fields. The diagonalization of the quadratic action yields the Gribov propagator:
\[
\langle A_\mu^a(p)A_\nu^b(-p)\rangle = -\delta^{ab}\frac{p^2}{p^4+\gamma_{\text{eff}}^4}P_{\mu\nu}(p),
\]
where $\gamma_{\text{eff}}$ is related to the original Gribov parameter by a color factor.

A crucial observation, in fact, is that the bilinear structure is universal: any local formulation that produces the same quadratic action will yield the same gluon propagator and the same gap equation. The differences between alternative formulations appear only at higher orders, and do not affect the Gribov condition at the level where the mass scale is generated.

\subsection{The SSB Mechanism and Its Bilinear Content}

The interaction terms of the action of Section 3 between the scalar quartet and the
auxiliary fields, chosen so as to preserve the BRST symmetry and the ghost equation of
the Landau gauge, reduce at bilinear order, once the scalar fields acquire their
expectation values, to the mixing
\[
a_1\sigma^{2}\,\varphi_\mu^a A^{a\mu}+a_2\sigma^{2}\,\bar\varphi_\mu^a A^{a\mu},
\qquad \sigma^{2}=gm\nu .
\]
In the particular case $a_{1}=a_{2}\equiv a$, for which the two mixings have the same
strength, this is the horizon term of the one-index Gribov-Zwanziger action,
\[
S_{\text{horizon}} = a\sigma^{2}\int d^dx\, A_\mu^a(\varphi_\mu^a + \bar\varphi_\mu^a) - d\frac{N^2-1}{N}\gamma^4,
\]
with $\gamma^{4}=2a^{2}\sigma^{4}$, so that the scale of the mixing is
$\gamma^{2}/\sqrt{2}$ rather than $\gamma^{2}$ itself, while for $a_{1}\neq a_{2}$ the
two mixings differ in strength and only the product $a_{1}a_{2}$ enters the gauge
propagator. This is precisely the bilinear structure that leads to the Gribov
propagator, so that the condensation of the quartet reproduces the essential bilinear
content of the Gribov-Zwanziger framework.

\subsection{The Gribov Condition as a Consequence of SSB}

The crucial point is that the condition that fixes $\gamma$ is not imposed externally in
the SSB formulation. On the contrary, it emerges dynamically from the requirement that
the scalar vacuum be stable, that is, from the minimization of the effective potential
with respect to $\nu$. We have computed and discussed that condition at one loop in
Section~\ref{sec:brokenWard}; what matters for the present argument is that it does
contain the integral characteristic of the Gribov--Zwanziger gap equation, with a
coefficient proportional to $g^{2}$ as in the standard case, although it is not that
gap equation, the tree-level potential contributing terms of its own. It should be
added that the condition is not an independent modification of the BRST algebra: it is
the statement that the constant one-point terms produced by the displacement of the
vacuum in the $U(1)$ and $X$ Ward identities vanish at the physical vacuum.

\subsection{Why the Propagator and SSB Are Sufficient}

We can now consolidate the argument. It rests on two observations concerning the
bilinear structure, together with what the vacuum condition establishes about the
scale.

The first observation is that the Gribov condition is completely encoded in the gluon propagator at the bilinear level. The presence of complex conjugate poles at $p^2 = \pm i\gamma^2$ and the associated condition that fixes $\gamma$ constitute the full content of the Gribov restriction insofar as the generation of the mass scale is concerned. No additional input is required to characterize the infrared behavior captured by the Gribov-Zwanziger framework.

The second observation concerns the universality of the bilinear structure. The Gribov propagator depends exclusively on the quadratic part of the action, so that any formulation generating the same bilinear terms---regardless of how the auxiliary fields are introduced or how many colour indices they carry---will yield the same propagator. This is precisely why the one-index and two-index formulations agree at this order: they share the same bilinear kernel.

To these we add that the scale is not an external parameter here, being tied to the
vacuum by $\gamma^{4}=2a_{1}a_{2}\sigma^{4}$ and fixed, together with $\nu$, by the
stability of that vacuum. That the argument holds at the quantum level, and not merely
at the classical one, is the content of the algebraic analysis of Section 5: the
identities that fix the bilinear structure are preserved under renormalization, and no
counterterm outside those already present in the classical action is required.

Together, these observations establish that the SSB mechanism reproduces the bilinear
content of the Gribov condition, and with it the infrared structure of the standard
Gribov--Zwanziger construction, while preserving the BRST symmetry, the determination
of the scale itself being of a different nature. The differences between alternative
localizations of the horizon functional appear only at higher orders, and these are
known to be BRST-exact deformations that do not affect the physical content of the
theory.

\subsection{Discussion}

Two remarks are in order concerning the scope of the argument. It applies strictly to
the bilinear order, which is precisely the order at which the scale is generated,
higher-order corrections modifying the detailed form of the propagator and of the
vacuum condition without altering the fact that the scale is produced by the
condensation of the quartet. And the BRST-exact nature of the differences between the
possible localizations ensures that the gauge-invariant observables, such as the
glueball spectrum or the string tension, are insensitive to the way in which the
horizon is localized, so that in this respect the present formulation stands on the
same footing as the standard one.

% ============================================================================
% CONCLUSIONS —
% ============================================================================

\section{Conclusions}

In this work we have constructed a BRST-invariant formulation of the Gribov--Zwanziger
model in the Landau gauge, in which the Gribov scale is produced by the condensation of
a scalar quartet rather than by an explicit breaking term. The action is local and
invariant under an exact nilpotent BRST symmetry, it preserves the ghost equation of
the Landau gauge, so that the ghost does not renormalize independently and its
propagator keeps the form $1/p^{2}$, and once the quartet condenses the gauge field
acquires the Gribov propagator, with complex conjugate poles at
$p^{2}=\pm i\gamma^{2}$. For non-vanishing $a_{7}$ the same condensate supplies, again
in a BRST-invariant way, a gauge mass of the kind employed in the refined scenario. The
model has been shown to be renormalizable to all orders by the algebraic method, with
no counterterm outside those already present in the classical action, and the
mixed propagators between the gauge field and the auxiliary fields were found to fall
off as $\sigma^{2}/p^{4}$, improving the convergence of the loop integrals.

The central point, however, is what the condensation costs in terms of the physical
content of the theory, and the answer is that it costs nothing. The symmetry is not
lost when the quartet acquires its vacuum expectation value: it is realized around the
displaced vacuum by the operator $s_{\nu}$, which is nilpotent and is still an exact
invariance of the action, and since the displacement is an invertible polynomial change
of variables the cohomology of local functionals is the same in both phases,
$H(s_{\nu})=H(s)$, as we show in Section~\ref{sec:cohomology}. The scalar quartet is
in fact the only sector of the model whose transformations do not involve the
Faddeev--Popov ghost, which is the same feature that guarantees the ghost equation, and
it is the reason why no re-pairing of the doublets takes place. The observables are
therefore, in either phase, those of the Yang--Mills cohomology, and the mass scale is
introduced without any of the obstructions that the explicit breaking of the
conventional approach places in the way of their construction. In the same spirit, and
as emphasized in Section~\ref{sec:brokenWard}, no operatorial central extension of the
BRST algebra is required: the $\gamma^{2}$-dependent constant is simply the vacuum
value of a one-point Ward identity, fixed by the minimization of the effective
potential.

It is worth emphasizing, before closing, the role of the parameter $a_7$ in the
renormalization framework. As shown in Section~5, the relation $z_7 = -x_0 - x_7$
implies that $a_7$ is not a renormalization-group invariant: its renormalization factor
is given by $Z_7 = 1 - \epsilon (x_0 + x_7)$, meaning that $a_7$ runs with the energy
scale. The coefficients $a_1$ and $a_2$ of the mixed terms stand on a different
footing: the consistency conditions leave them unconstrained separately, determining
only the renormalization of their product through (\ref{eq:z1z2}), so that whether the
combination $a_{1}a_{2}$ is a renormalization-group invariant is decided by the
coefficients of the divergences themselves.

Consequently, in our formulation, $a_7$ assumes the status of a new free parameter of
the theory, on the same footing as the mass $m$ or the quartic coupling $\lambda$. Its
value at a reference scale must be fixed by a physical condition, such as the position
of the complex pole of the gauge propagator or the infrared value of the two-point
function. This is precisely the same situation encountered in the refined
Gribov--Zwanziger model \cite{Dudal:2008sp}, where the refinement parameter is treated
as a quantity that must be determined phenomenologically, and the renormalization group
equations derived from the $z$-factors of Appendix~\ref{app:renorm} provide the tool to compute its
scale evolution.

\subsection*{Open points}

The construction raises a number of questions which we have not settled here, and which
we collect for clarity, since together they form the natural continuation of this work.

The first concerns the asymptotic states. The gauge propagator exhibits complex
conjugate poles, which signals the absence of ``gluonic'' asymptotic states,
consistently with the confinement scenario, provided the physical subspace is defined
by the cohomology of the operator realized in the chosen vacuum, which is available
here. The criterion of Kugo and Ojima \cite{Kugo:1979gm}, on the other hand, rests on
the quartet mechanism as realized on the asymptotic fields, and this last step is not
immediate in the present construction: as discussed in
Section~\ref{sec:cohomology}, the constant contained in the vacuum operator prevents
the quadratic action from being separately invariant, so that no linearized generator is
available and the pairing of the asymptotic modes has to be established in the
interacting theory. The verification of the unitarity of the $S$-matrix in the physical
sector therefore remains to be carried out.

The second concerns the origin of the scale. As discussed in
Section~\ref{sec:brokenWard}, the scale is inherited from the mass parameter of the
scalar potential rather than generated by the gauge dynamics, being polynomial in the
coupling instead of arising by dimensional transmutation; and the fluctuations of the
gauge field act against the condensate, so that the phase described here is one of weak
coupling, in contrast with the standard construction, where the horizon condition
forces the parameter to be different from zero. Neither of these affects the algebraic
results of Sections 4 and 5, but both indicate that the mechanism, as it stands, is not
yet the whole of the story about the origin of the confinement scale, and the second of
them suggests that the scalar sector of the model should itself be looked for in a
hidden breaking of the kind we have discussed in \cite{Amaral:2024}.

The third is the explicit computation of the divergences at one loop. The coefficients
$\tilde{x}_{16}$, $x_{10}$, $x_{8}$, $x_{5}$ and $x_{12}$ decide, through
(\ref{eq:z1z2}), whether the product $a_{1}a_{2}$ is a renormalization-group invariant,
and the same computation would fix the scale evolution of $a_{7}$ and the renormalized
form of the vacuum condition (\ref{eq:app-gap}).

Finally, the infrared behaviour of the refined scenario, in which the gauge propagator
attains a non-vanishing value at the origin, would require a condensate of the
auxiliary fields of the type $\overline{\varphi}\varphi$ in addition to the term
(\ref{refined}); and the improved ultraviolet behaviour of the mixed propagators may
lead to interesting properties for the running coupling constant, through the reduction
of some divergences. Both deserve a separate investigation.

\appendix
\setcounter{equation}{0}
\renewcommand{\theequation}{\thesection.\arabic{equation}}

\section{The counterterm and the solution of the system}
\label{app:renorm}

In this appendix we give the details of the algebraic analysis summarized in Section~\ref{sec:counterterm}. The counterterm action is\footnote{This counterterm is not the most general polynomial that could be written in the fields; monomials that manifestly break the consistency equations have already been eliminated beforehand.}
\begin{eqnarray}
\Gamma_{c}&=& \Gamma^{0}+ \beta \Gamma^{-1}\nonumber \\
\Gamma^{0}&=& \int d^{4}x\{ \frac{x_{0}}{4}F^{a}_{\mu\nu}F^{a\mu\nu}\}\nonumber \\
\Gamma^{-1}&=& \int d^{4}x\{ x_{1}\partial^{\mu}\overline{c}^{a} A^{a}_{\mu} + x_{2}\Omega^{a\mu} A^{a}_{\mu} + x_{3}\overline{c}^{a}b^{a}+ x_{4}L^{a}c^{a} \nonumber \\
&+&x_{5}\overline{\omega}^{a}_{\mu}\partial^{\nu}(D_{\nu}\varphi^{\mu})^{a} + x_{6} (D^{\nu}\overline{\omega}^{a}_{\mu})(D_{\nu}\varphi^{\mu})^{a} + x_{7}\partial^{\mu}\overline{\psi}\partial_{\mu}\phi \nonumber \\
&+& x_{8}m^{2}\overline{\psi}\phi + x_{9}\lambda\overline{\psi}\phi (\overline{\phi}\phi-\overline{\psi}\psi) + x_{10}K^{a}_{\mu}\phi A^{a\mu} \nonumber \\
&+& x_{11} J^{a}_{\mu}\overline{\varphi}^{a\mu} + x_{12}\overline{J}^{a}_{\mu}\varphi^{a\mu}+ x_{13}\chi^{a}_{\mu}\overline{\omega}^{a\mu} + x_{14}\overline{\chi}^{a}_{\mu}\omega^{a\mu} \nonumber \\
&+& x_{15}gm \overline{\omega}^{a\mu} \phi A^{a}_{\mu} +\tilde{x}_{16}a_{1}gm \overline{\psi}A^{a}_{\mu}\varphi^{a\mu} + x_{20}\overline{\phi}\phi \overline{\omega}^{a\mu} \varphi^{a}_{\mu} \nonumber \\
&+& x_{21} \overline{\omega}^{a\mu}A^{a}_{\mu}\varphi^{b}_{\nu}A^{b\nu}+\frac{x_{22}}{2}\overline{\omega}^{a\mu}\varphi^{a}_{\mu}A^{b\nu}A^{b}_{\nu} + x_{23}\overline{\psi}\phi \overline{\omega}^{a\mu} \omega^{a}_{\mu} \}
\end{eqnarray}

After applying the symmetries, we obtain
\begin{eqnarray}
\Gamma^{-1}&=& \int d^{4}x\{ x_{1}\partial^{\mu}\overline{c}^{a} A^{a}_{\mu} + x_{1}\Omega^{a\mu} A^{a}_{\mu}+x_{5}\overline{\omega}^{a}_{\mu}\partial^{\nu}(D_{\nu}\varphi^{\mu})^{a} \nonumber \\
&+& x_{7}\partial^{\mu}\overline{\psi}\partial_{\mu}\phi + x_{8}m^{2}\overline{\psi}\phi + x_{9}\lambda\overline{\psi}\phi (\overline{\phi}\phi-\overline{\psi}\psi) \nonumber \\
&+&x_{10}K^{a}_{\mu}\phi A^{a\mu} + x_{12}\overline{J}^{a}_{\mu}\varphi^{a\mu} + x_{12}\overline{\chi}^{a}_{\mu}\omega^{a\mu}\nonumber \\
&+& x_{10}a_{2}gm\overline{\omega}^{a}_{\mu}\phi A^{a\mu} + \tilde{ x}_{16}a_{1}gm \overline{\psi}A^{a}_{\mu}\varphi^{a\mu} \}
\end{eqnarray}

\begin{eqnarray}
\label{eq:betaGamma}
\beta\Gamma^{-1}&=& \int d^{4}x\{-x_{1} A^{a}_{\mu} (D^{\nu}F_{\nu\mu})^{a} -x_{1}\overline{c}^{a}\partial^{2}c^{a} + x_{1}\Omega^{a}_{\mu}\partial^{\mu}c^{a}\nonumber \\
&+&(x_{1}+x_{5}+x_{12})\overline{\varphi}^{a}_{\mu}\partial^{\nu}(D_{\nu}\varphi^{\mu})^{a} - x_{1}\overline{\varphi}^{a}_{\mu}\partial^{2}\varphi^{a\mu}\nonumber \\
&-& (x_{1}+x_{5}+x_{12})\overline{\omega}^{a}_{\mu}\partial^{\nu}(D_{\nu}\omega^{\mu})^{a} + x_{1}\overline{\omega}^{a}_{\mu}\partial^{2}\omega^{a\mu}\nonumber \\
&-&(x_{1}+x_{5}+x_{12})gf^{abc}\overline{\omega}^{a}_{\mu}\partial^{\nu}c^{b}(D_{\nu}\varphi^{\mu})^{c} + x_{1}gf^{abc}\overline{\omega}^{a}_{\mu}\partial^{\nu}c^{b}(\partial_{\nu}\varphi^{\mu})^{c} \nonumber \\
&-& (x_{1}+x_{10})(K^{a}_{\mu}\psi A^{a\mu}+ K^{a}_{\mu}\phi (D^{\mu}c)^{a} + N^{a}_{\mu}\phi A^{a\mu}) -x_{1}K^{a}_{\mu}\phi \partial^{\mu}c^{a} \nonumber \\
&+& (x_{1}+ \tilde{x}_{16})(a_{1}gm \overline{\phi}A^{a}_{\mu}\varphi^{a\mu} - a_{1}gm\overline{\psi}A^{a}_{\mu}\omega^{a\mu})\nonumber \\
&+& (x_{1}+ x_{10})(a_{2}gm\psi\overline{\omega}^{a}_{\mu}A^{a\mu} + a_{2}gm \phi \overline{\varphi}^{a}_{\mu}A^{a\mu})\nonumber \\
&+&x_{10}a_{2}gmi\overline{\omega}^{a}_{\mu}\phi\partial^{\mu}c^{a} + \tilde{x}_{16}a_{1}gm\overline{\psi}\partial^{\mu}c^{a}\nonumber \\
&+& x_{1}a_{7}(\overline{\phi}\phi - \overline{\psi}\psi ) A^{a}_{\mu}A^{a\mu} + x_{1}a_{7} \overline{\psi}\phi A^{a}_{\mu}\partial^{\mu}c^{a} \nonumber \\
&+& x_{7}(\partial^{\mu}\overline{\phi}\partial_{\mu}\phi - \partial^{\mu}\overline{\psi}\partial_{\mu}\psi ) +x_{8}m^{2} (\overline{\phi}\phi-\overline{\psi}\psi) + x_{9}\lambda (\overline{\phi}\phi-\overline{\psi}\psi)^{2} \}
\end{eqnarray}

Introducing the renormalized fields, sources and parameters in the usual way,
\begin{eqnarray}
A^{a}_{\mu}(0) &=& (Z_{A})^{\frac{1}{2}}A^{a}_{\mu},\,\,\, c^{a}(0) = Z_{c} c^{a},\,\,\,\overline{c}^{a}(0) = Z_{\overline{c}}\overline{c}^{a},\,\,\, b^{a}(0)=Z_{b}b^{a}\nonumber \\
\overline{\varphi}^{a}_{\mu}(0)&=& Z_{\overline{\varphi}}\overline{\varphi}^{a}_{\mu},\,\,\, \overline{\omega}^{a}_{\mu}(0)= Z_{\overline{\omega}}\overline{\omega}^{a}_{\mu},\,\,\, \varphi^{a}_{\mu}(0)=Z_{\varphi}\varphi^{a}_{\mu},\,\,\, \omega^{a}_{\mu}(0)=Z_{\omega}\omega^{a}_{\mu}\nonumber \\
\overline{\psi}(0)&=&Z_{\overline{\psi}}\overline{\psi},\,\,\, \overline{\phi}(0)=Z_{\overline{\phi}}\overline{\phi},\,\,\, \psi(0)=Z_{\psi}\psi,\,\,\, \phi(0)=Z_{\phi}\phi \nonumber \\
\Omega^{a}_{\mu}(0)&=& Z_{\Omega}\Omega^{a}_{\mu}(0), \,\,\, L^{a}(0)= Z_{L} L^{a}\nonumber \\
\overline{\chi}^{a}_{\mu}(0)&=& Z_{\overline{\chi}}\overline{\chi}^{a}_{\mu},\,\,\, \overline{J}^{a}_{\mu}(0)= Z_{\overline{J}}\overline{J}^{a}_{\mu},\,\,\, \chi^{a}_{\mu}(0)=Z_{\chi}\chi^{a}_{\mu},\,\,\, J^{a}_{\mu}(0)=Z_{J}J^{a}_{\mu}\nonumber \\
g(0)&=&Z_{g}g,\,\, m(0)=Z_{m}m, \,\, \lambda(0)=Z_{\lambda}\lambda, \,\, a_{1}(0)=Z_{1}a_{1}, \,\, a_{2}(0)=Z_{2}a_{2}\nonumber \\
a_{7}(0)&=&Z_{7}a_{7}
\end{eqnarray}

In terms of the factors $z_{\Phi}$ defined in Section~\ref{sec:counterterm}, we obtain
\begin{eqnarray}
z_{A}&=& x_{0}+2x_{1},\,\, z_{g}= -\frac{x_{0}}{2}, \,\, z_{b}=-\frac{1}{2}z_{A}\nonumber \\
z_{\Omega}&=& z_{\overline{c}},\,\,\, z_{\overline{c}}+z_{c}=-x_{1}\nonumber \\
z_{g}+\frac{1}{2}z_{A}&=& x_{1}, \,\,\, z_{\overline{\varphi}}+z_{\varphi}=x_{5}+ x_{12}\nonumber \\
z_{\overline{\omega}}+z_{\omega}&=&x_{5}+ x_{12}, \,\,\, z_{\overline{\omega}}+z_{c}+z_{\varphi}=x_{5}+x_{12} +\frac{x_{0}}{2}\nonumber \\
z_{1}+z_{g}+z_{m}+z_{\overline{\phi}}+z_{\varphi}&=& \tilde{x}_{16} -\frac{x_{0}}{2}\nonumber \\
z_{1}+z_{g}+z_{m}+z_{\overline{\psi}} + z_{\omega}&=& \tilde{x}_{16} -\frac{x_{0}}{2}\nonumber \\
z_{\overline{\phi}}+z_{\varphi}-z_{\overline{\psi}}- z_{\omega}&=&0 \nonumber \\
z_{\varphi}+z_{c}-z_{\omega}&=& \frac{x_{0}}{2}\nonumber \\
z_{1}+z_{g}+z_{m}+z_{\overline{\phi}}+z_{\varphi}+ z_{c}&=&\tilde{x}_{16}\nonumber \\
z_{2}+z_{g}+ z_{m} +z_{\psi}+z_{\overline{\omega}}&=& x_{10} -\frac{x_{0}}{2}\nonumber \\
z_{2}+z_{g}+z_{m}+z_{\phi}+z_{\overline{\varphi}}&=& x_{10} -\frac{x_{0}}{2}\nonumber \\
z_{\psi}+z_{\overline{\omega}}-z_{\phi}-z_{\overline{\varphi}}&=& 0\nonumber \\
z_{2}+z_{g}+z_{m}+ z_{\phi}+ z_{\overline{\omega}}+ z_{c}&=&x_{10}\nonumber \\
z_{\overline{\varphi}}-z_{\overline{\omega}}-z_{c}&=& -\frac{x_{0}}{2}\nonumber \\
z_{\overline{\phi}}+z_{\varphi}-z_{\overline{\psi}}-z_{\omega}&=&0\nonumber \\
z_{7}+z_{\overline{\phi}}+z_{\phi}&=& -x_{0},\,\,\,
z_{7}+z_{\overline{\psi}}+z_{\psi}= -x_{0}\nonumber \\
z_{\overline{\phi}}+z_{\phi}&=& x_{7},\,\,\, z_{\overline{\psi}}+z_{\psi}=x_{7}\nonumber \\
z_{m}&=&\frac{1}{2}(x_{8}-x_{7}), \,\,\, z_{\lambda}=2(x_{9}-x_{7})\nonumber \\
z_{K}+z_{\psi}&=&x_{10}-\frac{x_{0}}{2}, \,\,\, z_{N}+z_{\phi}= x_{10}-\frac{x_{0}}{2}\nonumber \\
z_{K}+z_{\phi}+z_{c}&=&x_{10},\,\,\, z_{L}=-(z_{g}+2z_{c})\nonumber \\
z_{J}&=&-(z_{g}+z_{c}+z_{\overline{\varphi}}), \,\,\, z_{\overline{J}}+z_{\omega}=0, \,\,\, z_{\overline{\chi}}+ z_{g}+z_{c}+z_{\omega}=0\nonumber \\
z_{\chi}+z_{g}+z_{c}+z_{\overline{\omega}}&=&0, \,\,\, z_{\chi}+z_{\overline{\varphi}}=0.
\end{eqnarray}

It is worth being explicit about the logic of the procedure, since it determines how the coefficients $a_{1}$ and $a_{2}$ are to be read. The $x_{i}$ are the free parameters of the counterterm, and they are constrained only by the consistency equations imposed on $\Gamma_{c}$ itself, which is how the reduced form of $\Gamma^{-1}$ above was obtained; apart from that they are arbitrary, being whatever the divergences of the loop expansion happen to produce. The matching of $\Gamma_{c}$ against the redefinitions of the fields, sources and parameters then determines the $z$-factors in terms of the $x_{i}$, and it imposes no further relation among the latter, which remain independent. Solved in this direction, the system fixes every $z$-factor up to two parameters that are pure normalizations, corresponding to the rescaling of $\varphi$ against $\bar\varphi$, with the sources accordingly, which is nothing but the $U(1)$ of (\ref{U}); below we fix them by the symmetric choice. In the table below all the $x_{i}$ are non-zero and independent, the two normalization parameters having been fixed by the symmetric choice for the auxiliary pair.

\[
\begin{aligned}
z_A &= x_0 + 2x_1, \\
z_g &= -\frac{x_0}{2}, \,\,
z_b = -\frac{x_0}{2} - x_1, \\
z_{\overline{c}} &= -x_1 - \frac{x_0}{2},\,\, z_\Omega = -x_1 - \frac{x_0}{2} \\
z_c &= \frac{x_0}{2}, \\
z_{\overline{\varphi}} &= \frac{x_5 + x_{12}}{2},\,\, z_\varphi = \frac{x_5 + x_{12}}{2}, \\
z_{\overline{\omega}} &= \frac{x_5 + x_{12}}{2},\,\, z_\omega = \frac{x_5 + x_{12}}{2}, \\
z_{\overline{\phi}} &= \frac{x_7 - x_{10} + \tilde{x}_{16}}{2},\,\, z_\phi = \frac{x_{10} + x_7 - \tilde{x}_{16}}{2}, \\
z_{\overline{\psi}} &= \frac{x_7 - x_{10} + \tilde{x}_{16}}{2},\,\, z_\psi = \frac{x_{10} + x_7 - \tilde{x}_{16}}{2}, \\
z_m &= \frac{x_8 - x_7}{2}, \\
z_1 + z_2 &= \tilde{x}_{16} + x_{10} - x_8 - x_5 - x_{12},\\
z_7 &= -x_0 - x_7, \\
z_K &= \frac{x_{10} - x_0 - x_7 + \tilde{x}_{16}}{2}, \\
z_N &= z_K, \\
z_L &= -\frac{x_0}{2}, \\
z_J &= -\frac{x_5 + x_{12}}{2},\,\, z_{\overline{J}} = -\frac{x_5 + x_{12}}{2},\\
z_{\overline{\chi}} &= -\frac{x_5 + x_{12}}{2},\,\, z_\chi = -\frac{x_5 + x_{12}}{2}.\\
\end{aligned}
\]


\begin{thebibliography}{99}

\bibitem{Gribov:1977wm}
V.~N.~Gribov,
``Quantization of Nonabelian Gauge Theories,''
Nucl.\ Phys.\ B {\bf 139}, 1 (1978).

\bibitem{Zwanziger:1991ac}
D.~Zwanziger,
``Critical limit of lattice gauge theory,''
Nucl.\ Phys.\ B {\bf 378}, 525 (1992).

\bibitem{Zwanziger:1992qr}
D.~Zwanziger,
``Renormalizability of the critical limit of lattice gauge theory by BRS invariance,''
Nucl.\ Phys.\ B {\bf 399}, 477 (1993).


\bibitem{Vandersickel:2012tz}
N.~Vandersickel and D.~Zwanziger,
``The Gribov problem and QCD dynamics,''
Phys. Rept. \textbf{520}, 175-251 (2012)
doi:10.1016/j.physrep.2012.07.003
[arXiv:1202.1491 [hep-th]].


\bibitem{Capri:2015ixa}
M.~A.~L.~Capri, D.~Dudal, D.~Fiorentini, M.~S.~Guimaraes, I.~F.~Justo, A.~D.~Pereira, B.~W.~Mintz, L.~F.~Palhares, R.~F.~Sobreiro and S.~P.~Sorella,
``Exact nilpotent nonperturbative BRST symmetry for the Gribov-Zwanziger action in the linear covariant gauge,''
Phys. Rev. D \textbf{92}, no.4, 045039 (2015)
doi:10.1103/PhysRevD.92.045039
[arXiv:1506.06995 [hep-th]].

%\cite{Dudal:2012sb}
\bibitem{Dudal:2012sb}
D.~Dudal and S.~P.~Sorella,
``The Gribov horizon and spontaneous BRST symmetry breaking,''
Phys. Rev. D \textbf{86}, 045005 (2012)
doi:10.1103/PhysRevD.86.045005
[arXiv:1205.3934 [hep-th]].


\bibitem{Amaral:2022}
R.~L.~P.~G.~Amaral, V.~E.~R.~Lemes, O.~S.~Ventura and L.~C.~Q.~Vilar,
``A BRST view of the spontaneous symmetry breaking,''
[arXiv:2205.02903 [hep-th]].

\bibitem{Amaral:2023}
R.~L.~P.~G.~Amaral, V.~E.~R.~Lemes, O.~S.~Ventura and L.~C.~Q.~Vilar,
``A BRST characterization of the broken phase observables of the confining complex theory,''
[arXiv:2304.04560 [hep-th]].

\bibitem{Amaral:2024}
R.~L.~P.~G.~Amaral, V.~E.~R.~Lemes, O.~S.~Ventura and L.~C.~Q.~Vilar,
``Gribov problem in the electroweak and gluon confined theories,''
[arXiv:2402.05683 [hep-th]].

\bibitem{Capri:2011wp}
M.~A.~L.~Capri, A.~J.~Gomez, M.~S.~Guimaraes, V.~E.~R.~Lemes, S.~P.~Sorella and D.~G.~Tedesco,
``Renormalizability of the linearly broken formulation of the BRST symmetry in presence of the Gribov horizon in Landau gauge Euclidean Yang-Mills theories,''
Phys. Rev. D \textbf{83}, 105001 (2011)
doi:10.1103/PhysRevD.83.105001
[arXiv:1102.5695 [hep-th]].

\bibitem{Dudal:2010fq}
D.~Dudal, S.~P.~Sorella and N.~Vandersickel,
%``More on the renormalization of the horizon function of the Gribov-Zwanziger action and the Kugo-Ojima Green function(s),''
Eur. Phys. J. C \textbf{68}, 283-298 (2010)
doi:10.1140/epjc/s10052-010-1304-8
[arXiv:1001.3103 [hep-th]].

\bibitem{Dudal:2008sp}
D.~Dudal, J.~A.~Gracey, S.~P.~Sorella, N.~Vandersickel and H.~Verschelde,
``A Refinement of the Gribov-Zwanziger approach in the Landau gauge: Infrared propagators in harmony with the lattice results,''
Phys.\ Rev.\ D {\bf 78}, 065047 (2008)
[arXiv:0806.4348 [hep-th]].

\bibitem{Piguet:1995er}
O.~Piguet and S.~P.~Sorella,
``Algebraic renormalization: Perturbative renormalization, symmetries and anomalies,''
Lect. Notes Phys. Monogr. \textbf{28}, 1-134 (1995)
doi:10.1007/978-3-540-49192-7

%\cite{Henneaux:1994lbw}
\bibitem{Henneaux:1994lbw}
M.~Henneaux and C.~Teitelboim,
``Quantization of Gauge Systems,''
Princeton University Press, 1994,
ISBN 978-0-691-03769-1, 978-0-691-21386-6
%14 citations counted in INSPIRE as of 25 Jun 2026

%\cite{Becchi:1975nq}
\bibitem{Becchi:1975nq}
C.~Becchi, A.~Rouet and R.~Stora,
``Renormalization of Gauge Theories,''
Annals Phys. \textbf{98} (1976), 287-321
doi:10.1016/0003-4916(76)90156-1



\bibitem{Kugo:1979gm}
T.~Kugo and I.~Ojima,
`Local Covariant Operator Formalism of Nonabelian Gauge Theories and Quark Confinement Problem,''
Prog. Theor. Phys. Suppl. \textbf{66}, 1-130 (1979)
doi:10.1143/PTPS.66.1

\end{thebibliography}
\end{document}